\documentclass[a4paper,11pt]{article}
\pdfoutput=1 

\usepackage{jcappub} 

\usepackage[T1]{fontenc} 

\usepackage{amsmath}
\usepackage{graphicx}
\usepackage{amssymb}
\usepackage{color}

\newcommand{\qoi}{\rho_{0,i}}

\newcommand{\qob}{\rho_{0,b}}

\newcommand{\doi}{\delta_{0,i}}
\newcommand{\doj}{\delta_{0,j}}

\newcommand{\dmi}{\delta_{m,i}}
\newcommand{\dmj}{\delta_{m,j}}

\newcommand{\dgi}{\delta_{g,i}}

\newcommand{\Do}{\Delta_{0}}
\newcommand{\Doi}{\Delta_{0,i}}

\newcommand{\Dm}{\Delta_{m}}
\newcommand{\Dmi}{\Delta_{m,i}}
\newcommand{\Dmj}{\Delta_{m,j}}
\newcommand{\Dg}{\Delta_{g}}
\newcommand{\Dgi}{\Delta_{g,i}}
\newcommand{\Dgj}{\Delta_{g,j}}

\newcommand{\co}{C_{0,ij}}
\newcommand{\ho}{H_{0,ij}}
\newcommand{\cm}{C_{m,ij}}

\newcommand{\Co}{\mathbf{C_0}}

\newcommand{\Ho}{\mathbf{H_0}}

\newcommand{\Cm}{\mathbf{C_m}}

\newcommand{\Cg}{\mathbf{C_g}}

\newcommand{\tiCo}{{C_0}^{-1}}
\newcommand{\tiHo}{{H_0}^{-1}}

\newcommand{\ico}{\tiCo_{ij}}
\newcommand{\iho}{\tiHo_{ij}}

\newcommand{\yi}{y_i}
\newcommand{\yj}{y_j}
\newcommand{\bh}{\hat{b}}
\newcommand{\bb}{\bar{b}}

\newcommand{\sh}{\hat{\sigma_8}}

\newcommand{\aic}{\mathbf{AIC}}
\newcommand{\bic}{\mathbf{BIC}}

\newcommand{\glrt}{\mathbf{GLRT}}

\subheader{Accepted for publication in JCAP - 16th February 2016}
\title{Joint constraints on galaxy bias and $\sigma_8$ through the N-pdf of the galaxy number density}

\author[a,b]{Pablo Arnalte-Mur,}
\author[c]{Patricio Vielva,}
\author[a,b]{Vicent J. Mart{\'\i}nez,}
\author[c]{Jos\'e L. Sanz,}
\author[d]{Enn Saar,}
\author[e]{and Silvestre Paredes}
\affiliation[a]{Observatori Astron\`omic de la Universitat de Val\`encia, C/Catedr\`atic Jos\'e Beltr\'an, 2, 46980 Paterna,  Val\`encia, Spain.}
\affiliation[b]{Departament d'Astronomia i Astrof\'{i}sica, Universitat de Val\`encia, C/Dr. Moliner 50, 46100 Burjassot,  Val\`encia, Spain.}
\affiliation[c]{Instituto de F{\'\i}sica de Cantabria (CSIC-UC), Avda. de Los Castros s/n, E-39005 - Santander, Spain.}
\affiliation[d]{Tartu Observatoorium, EE-61602, T\~oravere, Estonia.}
\affiliation[e]{Departamento de Matem\'atica Aplicada y Estad\'{\i}stica, Universidad Polit\'ecnica de Cartagena, C/Dr. Fleming s/n, 30203 Cartagena, Spain.}
\emailAdd{pablo.arnalte@uv.es}
\emailAdd{vielva@ifca.unican.es}
\emailAdd{martinez@uv.es}
\emailAdd{sanz@ifca.unican.es}
\emailAdd{saar@to.ee}
\emailAdd{silvestre.paredes@upct.es}

\abstract{
We present a full description of the N-probability density function of the galaxy number density fluctuations. This N-pdf is given in terms, on the one hand, of the
cold dark matter correlations and, on the other hand, of the galaxy bias parameter. The method relies on the assumption commonly adopted that the dark matter
density fluctuations follow a local non-linear transformation of the initial energy density perturbations. The N-pdf of the galaxy number density fluctuations allows
for an optimal estimation of the bias parameter (e.g., via maximum-likelihood estimation, or Bayesian inference if there exists any \emph{a priori} information on
the bias parameter), and of those parameters defining the dark matter correlations, in particular its amplitude ($\sigma_8$). It also provides the proper
framework to perform model selection between two competitive hypotheses. The parameters estimation
capabilities of the N-pdf are proved by SDSS-like simulations (both ideal log-normal simulations and mocks obtained from Las Damas simulations), showing
that our estimator is unbiased. We apply our formalism to the 7th release of the SDSS main sample (for a volume-limited subset with absolute magnitudes
$M_r \leq -20$). We obtain $\bh = 1.193 \pm 0.074$ and $\sh = 0.862 \pm 0.080$,  for galaxy number density fluctuations in cells of a size of $30h^{-1}$Mpc.
Different model selection criteria show that galaxy biasing is clearly favoured.
}


\begin{document}
\maketitle
\flushbottom

\section{Introduction}
\label{intro}

Observations of the distribution of the galaxies \citep{yor00a,col01a} and state-of-the-art N-body simulations \citep{spr05a,Klypin2011a} show that the large scale structure (LSS) of the universe forms a network of filaments, clusters and voids, mostly defined by the cold dark matter fluctuations. Galaxies trace the dark matter density field through the bias parameter, which links the evolution of the matter gravitational potential and the galaxy formation and distribution. Actually, determining the galaxy bias is not only useful to trace the dark matter structure, but also helps to understand the process of galaxy formation and distribution \citep{Swanson2008g}. 

Depending on the galaxy type that is under consideration, different aspects of the LSS network are probed. For instance, Luminous Red Galaxies (LRGs) will be associated with dark matter halos, with the most luminous ones corresponding to large clusters (typically formed at the crossing of several filaments); spiral galaxies will be, however, better tracers for the filaments of the network, where they appear in a larger proportion; QSOs will serve as tracers of the distant universe, where galaxies are still in the phase of accretion of matter to the inner black hole; etc.

LSS surveys as NVSS \citep{Condon1998a}, 2MASS \citep{Jarrett2004a}, SDSS \citep{yor00a} or VIPERS \citep{Garilli2014a} have provided very useful information (in a large wavelength range) about galaxy bias: e.g. \cite{Boughn2002a} for the NVSS, e.g. \cite{Francis2010a} for the 2MASS,  e.g. \cite{zeh10a} for the SDSS and e.g. \citep{DiPorto2014} for VIPERS.
In most of the literature, the bias has been estimated using the 3D-correlation function of galaxies \citep{fry93a,Gaztanaga1994a,Croft1999b}, counts-in-cells statistics \citep{sigad00,col01a,marinoni05,kovac11}, the projected correlation function \citep{nor02b,Zehavi2005b,zeh10a,arnalte14}, the bi-spectrum \citep{guo09,pol14} and higher order moments of the galaxy distribution \citep{Szapudi1998f,ver02a,Swanson2008g,McBride2011d}. The bias can also be inferred estimating the 1-point distribution function (pdf) from counts in cells, assuming a model for the mass pdf and measuring the galaxy over-density, see e.g., \cite{sigad00, DiPorto2014} and references therein. Related,  though not identical reasoning, has been used in this paper. In particular, as pointed out by Kitaura, Jasche and Metcalf in \cite{kitaura10}, it is possible to make a Bayesian matter density field reconstruction assuming a log-normal prior and modelling the galaxy distribution by a Poissonian process. The log-normal model is also adopted as a prior for the density field in inference models for large-scale structure as the one introduced by Jasche and Kitaura \cite{jasche10} by means of the Hamiltonian Monte Carlo (HMC) algorithm.  Other approaches to determine the galaxy bias are usually based on the use of multivariate probability distributions, typically Gaussian or lognormal distributions, which are well suited priors for Bayesian analyses (see e.g. \cite{jasche13, ata15, granett15, jasche15} and references therein).

Our method  relies on the use of the whole set of N-pdf of the galaxy number density fluctuations. This multivariate probability density function depends on the bias parameter and the correlations of the underlying cold dark matter fluctuations. When this N-pdf is seen as a function of the bias parameter, it represents, in fact, the likelihood of the data (i.e., the galaxy catalogue) given the galaxy bias parameter. Therefore, the N-pdf provides a full description of the statistical properties of the galaxy number density field, and allows one to derive interesting bias estimators as the maximum-likelihood bias, the mean bias, etc. In addition, it also gives the opportunity of performing model selection among different galaxy biasing scenarios. Finally, our approach provides a coherent scheme for introducing any available information on the bias parameter in the form of \emph{prior} probabilities, following the Bayes theorem.

The paper is organized as follows. 
In  section~\ref{sec:model} the model is presented, and it illustrates how to derive the N-pdf of the galaxy number density fluctuations (section~\ref{subsec:model}), as well as how to perform parameter estimation (section~\ref{subsec:est}) and model selection (section~\ref{subsec:modelsel}). 
In section~\ref{sec:data} we present the real and simulated data used in this work: the real galaxy catalogues from the SDSS (section~\ref{subsec:sdss_data}), a set of lognormal simulations following our model (section~\ref{subsec:logn_simus}), and a set of mocks obtained from N-body cosmological simulations (section~\ref{subsec:lasdamas_simus}). 
We test the performance of our approach by analysing both sets of simulations in section~\ref{sec:tests}.
The results obtained from the application of the method to SDSS data are given in  section~\ref{sec:sdss}. 
Finally, we give our conclusions in  section~\ref{sec:final}.

Except when noted otherwise, we use a fiducial flat $\Lambda$CDM cosmological model with parameters $\Omega_m = 0.308$, $\Omega_{\Lambda} = 0.692$, $\sigma_8 = 0.8149$ based on the \emph{Planck-2015} results \cite{PlanckCollaboration2015b}. All the distances used are comoving and are given in terms of the Hubble parameter $h = H_0 / 100 \, \mathrm{km} \mathrm{s}^{-1} \mathrm{Mpc}^{-1}$.

\section{The galaxy number density model}
\label{sec:model}

As outlined in the Introduction, we aim to determine the N-pdf of the galaxy number density field, given the correlation properties of the dark matter density field, as well as the bias parameter $b$, that relates the galaxy formation with the fluctuations of the dark matter density field.

The description of the galaxy density field via its N-pdf provides a powerful framework to derive detailed statistical analyses. In particular, the N-pdf allows for a maximum-likelihood estimation (or Bayesian estimation, in the case of introducing any possible prior knowledge) of the parameters describing the galaxy distribution. In addition, model selection approaches can be also applied. 

These two approaches have been applied by \cite{vie08a,Vielva2010b} to the N-pdf of a local Gaussian deviation of the cosmic microwave background temperature fluctuations. This deviation was given in terms of a non-linear perturbative expansion of a Gaussian field. In the Sachs-Wolfe regime, and under certain conditions, this perturbative model defaults on the weak non-linear coupling inflationary model \citep[e.g.,][]{Komatsu2001c}.

\subsection{Distribution of the galaxy number density fluctuations}
\label{subsec:model}

Let us denote by $\qoi$, the initial energy density at a given position $i$, and by $\qob$ the mean initial energy density or the \emph{background} initial energy density. Then, the \emph{initial energy density fluctuation} at position $i$ is nothing but:
\begin{equation}
\label{eq:doi}
\doi = \frac{\qoi - \qob}{\qob},
\end{equation}
and the \emph{initial energy density contrast} is defined as:
\begin{equation}
\label{eq:Doi}
\Doi = 1 + \doi = \frac{\qoi}{\qob}.
\end{equation}
Trivially, $\left< \doi \right> \equiv 0$ and $\left< \Doi \right> \equiv 1$. As predicted by the standard inflation scenario \citep[e.g.][]{Liddle2000c}, the probability density function of the initial energy density fluctuations follows a multivariate normal distribution:
\begin{equation}
\label{eq:pdfo}
f\left( \doi \right) = \frac{1}{\left( 2\pi \right)^{N/2}\left| \Co \right|^{1/2}}
\exp{\left(-\frac{1}{2}\doi \ico \doj\right)},
\end{equation}
where $N$ represents the number of positions (assuming a given discretization or pixelization of space) in which the field is realized, and the $\Co$ matrix provides the Gaussian field correlations (i.e., $\left< \doi \doj \right> \equiv \co$).  Hereinafter, addition over repeated indices is assumed, i.e., 
$\doi \ico \doj \equiv \sum_{i=1}^N\sum_{j=1}^N \doi \doj \ico$. 

N-body simulations have been showing, since the mid 80s, that the dark matter density field shows a non-Gaussian behaviour, even when the seeds of the initial energy density perturbations are Gaussian \citep[e.g.,][]{Saslaw1985c,Suto1990b,Lahav1993b,Ueda1996d}. This non-Gaussianity is a reflection of the non-linear nature of the gravitational instability. This hint provided by N-body simulations is also confirmed by large-scale surveys, that also demonstrate that the galaxy number density follows a non-Gaussian distribution. Among all the local non-linear transformations, the log-normal model \cite[e.g.,][]{col91a,kay01a} provides an excellent description for the galaxy distribution, at least for the lowest non-linear orders. This model also provides an excellent framework to derive analytical expressions, and it is the one adopted in this work. It is convenient to express the matter density contrast $\Dmi$ as a function of the initial energy density fluctuations through the following transformation:
\begin{equation}
\label{eq:o2m}
\Dmi = \exp{\left(\alpha \frac{\doi}{\sigma_{\Do}}-\frac{\alpha^2}{2}\right)},
\end{equation}
where $\alpha$ is a constant and $\sigma_{\Do}^2 \equiv C_{0,ii} \equiv \left< \doi^2 \right>$. Conversely, the local inverse transformation is given by:
\begin{equation}
\label{eq:m2o}
\doi = \frac{\sigma_{\Do}}{\alpha} \left(\log \Dmi + \frac{\alpha^2}{2}\right).
\end{equation}
Attending to these definitions, it is straightforward to show that:
\begin{eqnarray}
\label{eq:relations1}
\alpha^2 & = & \log\left( 1 + \sigma_{\Dm}^2 \right) \\
\left< \Dmi \right> & = & 1 \nonumber \\
\left< \Dmi \Dmj \right> & = & \exp \left[ \left( \frac{\alpha}{\sigma_{\Do}}\right)^2 \co \right] \nonumber \\
\ho & = & \log \left( 1 + \cm \right) \nonumber, 
\end{eqnarray}
where $\sigma_{\Dm}^2 \equiv C_{m,ii}$, and $\ho \equiv \left(\frac{\alpha}{\sigma_{\Do}}\right)^2 \co$. Given equations~(\ref{eq:pdfo}) and~(\ref{eq:m2o}), it is trivial to compute the N-pdf associated with the cold dark matter density field:
\begin{equation}
\label{eq:pdfm}
f\left( \Dmi \right) = \frac{1}{\left( 2\pi \right)^{N/2}\left| \Ho \right|^{1/2}}
\frac{\exp{\left(-\frac{1}{2}\yi \iho \yj \right)}}{\Pi_{j=1}^N \Dmj},
\end{equation}
where we have defined the auxiliary parameters $\yi$ as
\begin{equation}
\label{eq:yidef}
\yi \equiv \log \left( \Dmi \sqrt{1 + \sigma_{\Dm}^2} \right) .
\end{equation}
The galaxy number density fluctuations ($\dgi$) are assumed to be related to the cold dark matter density through a local transformation in terms \citep{dek99,cac12} of the \emph{galaxy bias} $b$:
\begin{equation}
\label{eq:bias}
\dgi = b\dmi ~ \Longrightarrow ~ \Dmi = \frac{\Dgi + b - 1}{b},
\end{equation}
where $\Dgi$ is the galaxy number density contrast. Given this relation, and taking into account equation~(\ref{eq:pdfm}), it is straightforward to compute the N-pdf associated with the galaxy number density field:
\begin{equation}
\label{eq:pdfg}
f\left( \Dgi \right) = \frac{1}{\left( 2\pi \right)^{N/2}\left| \Ho \right|^{1/2}}
\frac{ \exp{ \left( -\frac{1}{2}\yi \iho \yj \right) }}{\Pi_{j=1}^N \left( \Dgj + b - 1 \right) },
\end{equation}
where we can use equations~(\ref{eq:yidef}) and (\ref{eq:bias}) to express $\yi$ in terms of the galaxy number density fluctuations ($\dgi$) and their correlations ($\Cg = b^2 \Cm$):
\begin{equation}
\label{eq:relations2}
\yi  =  \log{\left[ \frac{\sqrt{b^2 + \sigma_{\Dg}^2}}{b^2} \left( \Dgi + b - 1 \right) \right]} 
\end{equation}
where $\sigma_{\Dg}^2  \equiv C_{g,ii} = b^2 C_{m,ii}$.

\subsection{Parameter estimation}
\label{subsec:est}

In the context of parameter estimation, the N-pdf given by equation~(\ref{eq:pdfg}) can be seen as the likelihood function of observations (the values of the galaxy number density contrast $\Dgi$ at $N$ positions in space) given a galaxy clustering model.
This clustering model has two parts: on one hand the galaxy bias parameter $b$, and on the other the covariance of the matter fluctuations $\Cm$ fixed by the cosmological model.
These covariances are defined as $\cm = \left\langle \dmi \dmj \right\rangle$, and are therefore equal to
\begin{equation}
\cm = \xi_m(r_{ij}) ,
\end{equation}
where $r_{ij}$ is the comoving distance between the centers of the cells $i,j$, and $\xi_m(r)$ is the matter correlation function predicted by the cosmological model, filtered to account for the finite size of the cells. In our case, given a model we obtain the matter power spectrum $P_m(k)$ using the \textsc{Camb}\footnote{http://camb.info} software \cite{lew00a}, and calculate the corresponding $\xi_m(r)$ via a Fourier transform using \textsc{FFTLog} \cite{Hamilton2000a}.

In the simplest case we can assume that the cosmological model is known, fixing $\Cm$, so the only free parameter is the bias $b$. In this case, we can express the likelihood as
\begin{equation}
\label{eq:like}
L \left( \Dgi | b\right) \equiv f\left( \Dgi \right).
\end{equation}
The advantages of exploiting the N-pdf for statistical analyses, as parameter estimation or model selection, are clear. 
In particular, given the $\Dgi$ observed by a galaxy survey and the $\Cm$ from the assumed cosmological model, we can obtain the maximum-likelihood estimate of galaxy the bias $\bh$, as well as alternative estimates like the mean value.
In addition, if any \emph{a priori} information on the bias parameter $p\left(b\right)$ is available, it can be used together with equation~(\ref{eq:pdfg}) to provide the posterior probability of the bias parameter given the observations $p\left(b | \Dgi\right)$, following the Bayes' theorem:
\begin{equation}
\label{eq:post1}
p\left( b| \Dgi \right) \propto L\left( \Dgi | b \right)~p\left(b\right).
\end{equation}
This posterior probability allows performing a full Bayesian parameter estimation, as well as Bayesian model selection (e.g., Bayesian evidence).

In addition to the bias, we can also use this likelihood approach to constrain the parameters of the cosmological model via the covariance matrix for matter fluctuations, $\Cm$. In particular, we focus here on constraining the amplitude of the matter power spectrum, jointly with the galaxy bias $b$. We parameterize this amplitude using the standard parameter $\sigma_8$.\footnote{$\sigma_8^2$ corresponds to the matter density variance in spheres of radius $R = 8 h^{-1}\, \mathrm{Mpc}$, when using a linear model extrapolated to $z=0$.}
In order to introduce this additional parameter in equation~(\ref{eq:pdfg}), we first assume a fiducial value $\sigma_8^{\rm fid}$ and compute the corresponding covariance matrix $\mathbf{C_m}^{\rm fid}$. The covariance matrix then depends on $\sigma_8$ as
\begin{equation}
\label{eq:cm_sigma}
\mathbf{C_m} = \left( \frac{\sigma_8}{\sigma_8^{\rm fid}}\right)^2 \mathbf{C_m}^{\rm fid} .
\end{equation}
Introducing this expression for $\Cm$ into equation~(\ref{eq:pdfg}), we can then interpret $f(\Dgi)$ as the likelihood of the data $\Dgi$ given the two parameters $(b, \sigma_8)$,
\begin{equation}
\label{eq:like_bsigma}
L \left( \Dgi | b, \sigma_8 \right) \equiv f\left( \Dgi \right).
\end{equation}
In the same way as in the single-parameter case, this likelihood function can be used to obtain the joint maximum likelihood estimates for the two parameters $\bh$, $\sh$ or other estimates. We can also combine this likelihood with any prior on the parameters to obtain the joint posterior probability distribution,
\begin{equation}
\label{eq:post_bsigma}
p\left( b, \sigma_8 | \Dgi \right) \propto L\left( \Dgi | b, \sigma_8 \right)~p\left(b, \sigma_8 \right).
\end{equation}

In this work we always use flat wide priors either on $b$ (for equation~\ref{eq:post1}) or on $(b, \sigma_8)$ (for equation~\ref{eq:post_bsigma}), so these relations are just a \emph{normalization} of the posterior distributions. However, these prior probabilities could be used, e.g., to add information coming from independent observations constraining $\sigma_8$.

In principle, these two parameters could be degenerate, as they are both related to the overall galaxy clustering amplitude. 
This is the case in the analyses of galaxy bias based on two-point statistics, where the estimates of $b$ are completely degenerate with $\sigma_8$, and in fact one can only constrain the quantity $b \sigma_8$.
When using the N-pdf, however, we also have information on the shape of the distribution that breaks this degeneracy. 
Our model predicts a multivariate log-normal N-pdf for the matter density contrast $\Dmi$, (eq.~\ref{eq:pdfm}), but a different distribution for the galaxy density contrast $\Dgi$ (eq.~\ref{eq:pdfg}) when $b \neq 1$.
Therefore, variations in $\sigma_8$ change the overall variance of the distribution, while keeping the log-normal shape, while variations in $b$ change both the overall variance and the shape of the distribution, breaking the degeneracy.

One could also use this approach to constrain other cosmological parameters that affect the model covariance matrix $\Cm$, such as $\Omega_{m}$. However, these parameters affect as well the estimation of the galaxy density $\Dgi$ from the data, via the conversion from galaxy redshifts to distances. This means that the simple likelihood interpretation presented here is not valid in this case. For this reason, we keep all cosmological parameters except $\sigma_8$ fixed in our analysis. In section~\ref{subsec:lasdamas_results} we test the reliability of the method to this assumption.

\subsubsection{Estimation of the uncertainty on the parameters}
\label{subsubsec:uncertainty}

As described above, we can use the N-pdf of the galaxy density field to derive the maximum-likelihood estimation of the parameters $b$, $\sigma_8$. We estimate the uncertainty on these parameters using the Fisher matrix ($\mathbf{F}$) formalism.

In the case in which we keep $\sigma_8$ fixed and only fit for the bias $b$, the uncertainty $\sigma_{\bh}$ on our estimate $\bh$ is given by
\begin{equation}
\label{eq:post}
\sigma_{\bh}^2 = \mathbf{F_{\bh}}^{-1} ~;~~~ 
\mathbf{F_{\bh}} = -\left[\frac{\partial^2 \ell_b}{\partial b^2}\right]_{b\equiv\bh} ,
\end{equation}
where $\ell_b = \log L\left(\Dgi | b\right)$.
In the case in which we estimate both $b$ and $\sigma_8$ from our method, the covariance matrix $\mathbf{C}_{\bh, \sh}$ of the estimated parameters $\bh$, $\sh$ is given by
\begin{equation}
\mathbf{C}_{\bh, \sh} = \mathbf{F_{\bh, \sh}}^{-1} \, ; \,
\mathbf{F_{\bh, \sh}}= - \left[
\begin{array}{cc}
\displaystyle
\frac{\partial^2 \ell}{\partial b^2} &  \displaystyle \frac{\partial^2 \ell}{\partial b \partial \sigma_8} \\
\displaystyle \frac{\partial^2 \ell}{\partial \sigma_8 \partial b} & \displaystyle \frac{\partial^2 \ell}{\partial \sigma_8^2}
\end{array} \right]_{(b, \sigma_8) \equiv (\bh, \sh)} \, ,
\end{equation}
where in this case $\ell = \log L\left(\Dgi | b, \sigma_8\right)$. The diagonal terms of this matrix correspond to the variances of the estimated parameters
\begin{equation}
\sigma_{\bh} = \sqrt{C_{11}} \, ; \, \sigma_{\sh} = \sqrt{C_{22}} \, ,
\end{equation}
while the diagonal term corresponds to the covariance between the two parameters.

\subsection{Model selection}
\label{subsec:modelsel}

Even if \emph{a priori} information on $b$ is lacking, model selection can also be performed following a likelihood-based approach. 
In the context of the N-pdf of the galaxy density field, it is interesting to compare our full two-parameter model (defined by equations~\ref{eq:cm_sigma} and \ref{eq:like_bsigma}) to an alternative no bias scenario (i.e., $b \equiv 1$) in which we only fit for $\sigma_8$.
In this work, we consider the \emph{Akaike information criterion} \citep[AIC,][]{Akaike1973a}, the \emph{Bayesian information criterion} \citep[BIC,][]{schwarz1978a}, 
and the \emph{generalized likelihood ratio test} (GLRT). 
These criteria have been already applied (and explained in detail) in the context of a N-pdf derived for a non-Gaussian model describing a local transformation \citep{vie08a}, that could represent the weak non-linear coupling inflationary model \citep{Komatsu2001c}, in the Sachs-Wolfe regime. 
Given a certain hypothesis $H_\alpha$, defined both by the maximum-likelihood parameters $\theta_\alpha$\footnote{in our case, depending of the model, $\theta = b$ or $\theta = (b, \sigma_8)$} and a maximum log-likelihood value  $\ell_\alpha \equiv \log{L \left( \Dgi | \theta_\alpha\right)}$, the AIC and BIC are given by:
\begin{eqnarray}
\label{eq:select}
\aic\left(H_\alpha\right) & = & 2\left(N_p-\ell_\alpha\right)\\
\bic\left(H_\alpha\right) & = & 2\left(\frac{N_p}{2}\log{N}-\ell_\alpha\right) 
\end{eqnarray}
where $N_p$ is the number of free parameters in the model (in our case, either $N_p = 1$ or $N_p=2$), and $N$ is the number of data points (in our case, the number of grid points in which we computed $\Dg$). 
Therefore, for instance, a given hypothesis $H_\alpha$ is said to be favoured by the $\aic$ with respect to an alternative $H_\beta$, if $\aic\left(H_\alpha\right) < \aic\left(H_\beta\right)$. The same is applied to the $\bic$ model selection procedure.
Regarding the GLRT approach, $H_\alpha$ is said to be favoured over $H_\beta$, at a $\nu$ level (with $\nu > 0$), if $\log\left(\frac{L_{\alpha}}{L_{\beta}}\right) \geq \nu$.

\section{Data}
\label{sec:data}

In this section, we present the different data sets used in this work. In section~\ref{subsec:sdss_data} we explain the selection of our sample from the SDSS, and how we estimate the galaxy density field from it. Section~\ref{subsec:logn_simus} describes how we generated a set of log-normal simulations to test the consistency of our method. Finally, in section~\ref{subsec:lasdamas_simus} we present the Las Damas set of mocks which we used to test the method in realistic simulations.

\subsection{SDSS Main galaxy sample}
\label{subsec:sdss_data}

We use the method described in section~\ref{sec:model} to study the galaxy bias and matter power spectrum amplitude for a galaxy sample drawn from the 7th data release \citep{aba08a} of the SDSS main catalogue. 
We used the data provided by the New York University Value Added Catalogue \citep[NYU-VAGC,][]{bla05b}. In order to avoid problems due to the radial selection function, we selected a volume limited sample with \mbox{$M_r \leq -20$} in the redshift range $0.033 < z < 0.106$, where $M_r$ is the $K+E$ corrected $r$-band absolute magnitude. 
We chose this redshift range to ensure that the sample is volume limited, and also to use a sample consistent with the correlation function analysis of \cite{zeh10a} and the Las Damas mocks (see section~\ref{subsec:lasdamas_simus}).
We convert the measured angles and redshifts into co-moving coordinates using our fiducial cosmological model based on the \emph{Planck-2015} results.

In order to study the N-pdf as described in section~\ref{sec:model}, we need to estimate from this galaxy sample the galaxy overdensity field $\Dg$. We chose to study this field using a grid of cubic cells covering the volume of the survey.
In order to choose the physical size of the cells, we studied the variance between cells obtained at different resolutions. 
Comparing to the variance obtained for Poisson catalogues of the same volume and density, we can quantify the relative effect of shot noise.
Using this approach, we decided to adopt cubic cells with a side of $30\,h^{-1}\,\mathrm{Mpc}$, for which the observed variance is $\sim 7$ times larger than the Poisson noise, so we can assume that we are in the signal-dominated regime. 

Once our grid of cells is defined, we proceed as follows to estimate $\Dg$. In the first place, we compute the completeness for each of the cells ($c_i$) as the combination of two components: the radial and the angular selection functions. For the former we assume a constant selection in the redshift range mentioned above, and for the latter we use the angular mask from the NYU-VAGC in the \textsc{Mangle} software format \cite{ham04a,swa08a}. To simplify the selection function, we consider only the North Cap of the SDSS area. We use in our analysis only cells with a completeness $c_i \geq 0.8$. This leaves us with $N=582$ valid cells, with a volume of $V = 15.7\times10^6 \, h^{-3}\, \mathrm{Mpc}^3$, embedded in a box of $240\times480\times240 \, h^{-3}\, \mathrm{Mpc}^3$. The total number of galaxies in the selected cells is $N_g = 90,634$.

\begin{figure}
\begin{center}
\includegraphics[width=10.5cm,keepaspectratio]{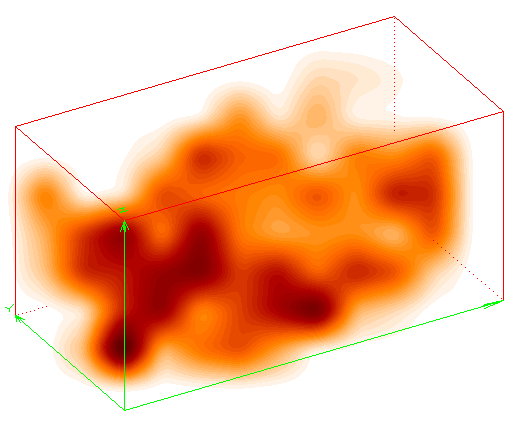}
\begin{center}
\includegraphics[width=7.4cm,keepaspectratio]{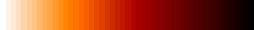}
\end{center}
\caption{\label{fig:data_map} 3D projection of the galaxy number density field corresponding to the SDSS catalogue used in this work. The colour palette used in the projection is shown at the bottom and corresponds to densities $1\leq \Delta_g \leq 2$ from left to right. }
\end{center}
\end{figure}

We obtain the number of cells $n_i$ in each of these accepted cells, and estimate the galaxy number density for each cell as
\begin{equation}
\rho_{g,i} = \frac{n_i}{c_i V_c} \, 
\end{equation}
where $V_c = (30 \, h^{-1} \, \mathrm{Mpc})^3$ is the volume of a cell. We finally compute the galaxy density contrast $\Dgi$ normalising $\rho_{g,i}$ by the mean galaxy density (eq.~\ref{eq:Doi}). 
In Figure~\ref{fig:data_map} we show a 3D projection of this galaxy density field obtained from the SDSS data.
The galaxy density contrast field $\Dgi$ in the $N$ cells estimated in this way is the quantity whose pdf is given by equation~(\ref{eq:pdfg}). As we already take into account the selection function here to define the `valid' cells, and to estimate the density, the method described in section~\ref{sec:model} can be applied directly. We only have to take into account the positions of the selected cells to calculate the model covariance matrix $\Cm$. 

\subsection{Lognormal simulations}
\label{subsec:logn_simus}

In order to test the method, we created a set of 100 lognormal realizations of the matter density field. These simulations are created following the same model described in section~\ref{sec:model}, and with the same survey geometry as the SDSS data (section~\ref{subsec:sdss_data}), so they can be used to assess the consistency of our method.
The input matter power spectrum $P_m(k)$ for these simulations corresponds to the best-fit model to the temperature + lensing \emph{Planck-2015} data 
\cite[][table 4, column 2]{PlanckCollaboration2015b}, computed using the \textsc{Camb} software.

We generate the lognormal simulations using the method described in \cite{lab10a} (which is equivalent to the one in \cite{Greiner2015a}), which is based on the fact that a lognormal random field is a local transformation of a Gaussian field. We compute the matter correlation function $\xi_m(r)$ via a Fourier transform of $P_m(k)$. We obtain the correlation function for the initial Gaussian field $\xi_0(r)$ using the transformation in equation~(\ref{eq:relations1}), as
\begin{equation}
\xi_0(r) = \left(\frac{\sigma_{\Do}}{\alpha}\right)^2 \log\left(1 + \xi_m(r)\right) ,
\end{equation}
and the corresponding power spectrum $P_0(k)$ as its inverse Fourier transform.
We then generate the Gaussian random field $\doi$ in a cubic grid using the standard method of generating Gaussian Fourier modes with variances given by $P_0(k)$ and then performing a Fast Fourier Transform (see e.g. section~7.4.1 of \cite{mar02a}). Finally, we obtain the corresponding matter density field $\Dmi$ using the transformation given by equation~(\ref{eq:o2m}).

To avoid possible boundary problems, we initially generate each of these lognormal simulations in a box of $(1440 \, h^{-1}\, \mathrm{Mpc})^3$, with cells of side $30\,h^{-1}\,\mathrm{Mpc}$. We then use the cell completeness defined for the SDSS data to define the same geometry with $N=582$ valid cells, and we keep only the value of the density field $\Dmi$ in those cells. In this way, we ensure that the grid points we use are always at a distance $\geq 480\,h^{-1}\,\mathrm{Mpc}$ from any of the borders of the box.

\begin{figure*}
\includegraphics[width=7.4cm,keepaspectratio]{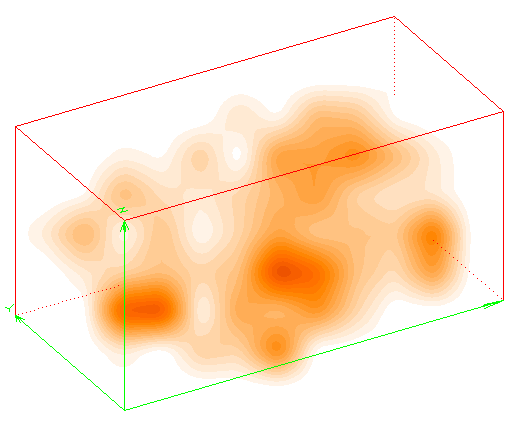}
\includegraphics[width=7.4cm,keepaspectratio]{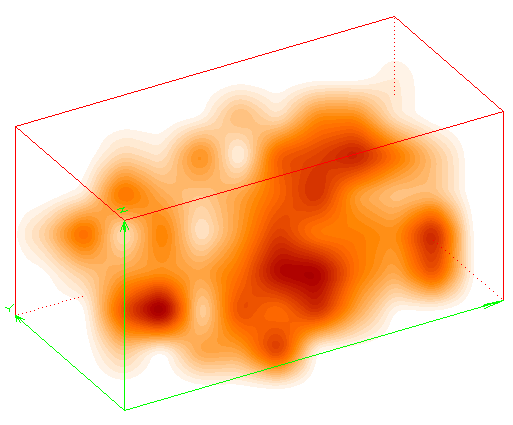}
\includegraphics[width=7.4cm,keepaspectratio]{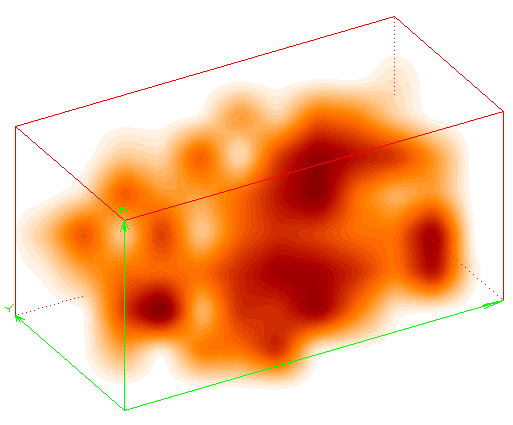}
\hspace{0.5cm}
\includegraphics[width=7.4cm,keepaspectratio]{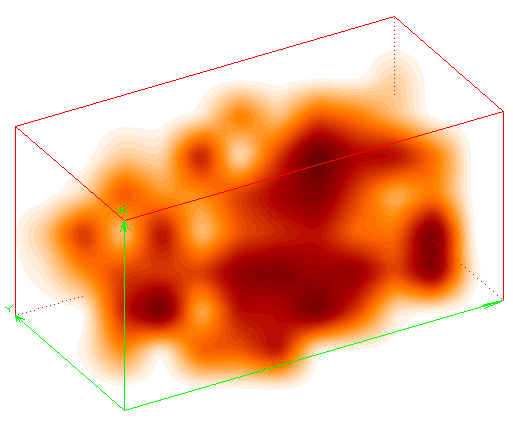}
\begin{center}
\includegraphics[width=7.4cm,keepaspectratio]{./escala1.png}
\end{center}
\caption{\label{fig:sims_maps} Galaxy number density field, shown as a 3D projection, for one of the realizations of the lognormal simulations. The underlying cold dark matter density field is the same for all the cases, but the galaxy bias is different in each of the panels. Bias values are, from left to right and top to bottom, $b = 0.5, 1.0, 1.5, 2.0$. The colour palette used in the projection is the same as in Figure~\ref{fig:data_map}. Notice how clustering increases as bias grows.}
\end{figure*}

For each of our 100 simulations of the matter density field $\Dmi$, given a value for the galaxy bias $b$, we can generate the corresponding galaxy density field $\Dgi$ using equation~(\ref{eq:bias}). In section~\ref{subsec:logn_results} we will explore the results we obtain for four input values of the bias: $b = 0.5, 1.0, 1.5, 2.0$. In Figure~\ref{fig:sims_maps} we represent 3D-projections of the galaxy number density field for these bias values obtained from one of our simulations (therefore the underlying realization of matter fluctuations is  the same for the 4 cases shown). It can be seen that, as it is expected, the clustering increases when the bias parameter is getting larger. We can compare this projection to that of the real data shown in Figure~\ref{fig:data_map}.

\subsection{Las Damas set of mock catalogues}
\label{subsec:lasdamas_simus}

As a further assessment of the N-pdf method, it is important to test it in more realistic simulated catalogues, which reproduce the observed distribution of galaxies. We chose to use the set of galaxy mocks obtained from the Las Damas simulations\footnote{http://lss.phy.vanderbilt.edu/lasdamas/} \cite{McBride2009a, kaz09a}, and in particular the \emph{gamma} release.
These mock catalogues were created matching both the selection effects and the clustering properties of the SDSS-DR7 real catalogues, and are therefore optimal for our purposes.
We use the set of mocks corresponding to the galaxy selection $M_r \leq -20$, to match the SDSS galaxy catalogue described in section~\ref{subsec:sdss_data}.

The mocks used here were obtained from the Esmeralda dark matter simulation: a set of 30 N-body realizations, each containing $1250^3$ particles in a volume of $(640\, h^{-1}\, \mathrm{Mpc})^3$. 
These realizations are created using a standard $\Lambda$CDM model with parameters $\Omega_m = 0.25$, $\Omega_{\Lambda} = 0.75$, $\sigma_8 = 0.8$.
We use this same cosmological model when analysing the Las Damas mocks.

The simulations are populated by galaxies using the halo occupation distribution (HOD) formalism, with the HOD parameters tuned to reproduce the observed number density and projected correlation function $w_p(r_p)$ (at scales $r_p \in [0.3, 30]\, h^{-1} \, \mathrm{Mpc}$, as studied in \cite{zeh10a}) of the corresponding SDSS catalogues. 
Finally, the mock galaxy catalogues are obtained by applying the SDSS angular selection mask from NYU-VAGC and appropriate redshift limits. Both selections match those we used to select the real SDSS catalogue presented in section~\ref{subsec:sdss_data}.
We use the \emph{North-only} version of the mocks for which four non-overlapping mocks are obtained from each simulation box. Hence, we end up with a total of 120 mock galaxy catalogues.

\begin{figure}
\begin{center}
\includegraphics[width=10.5cm,keepaspectratio]{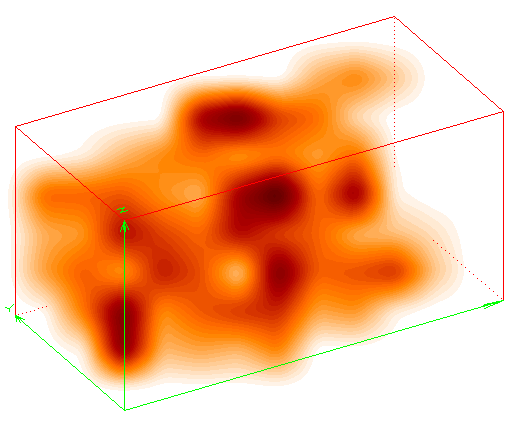}
\begin{center}
\includegraphics[width=7.4cm,keepaspectratio]{./escala1.png}
\end{center}
\caption{\label{fig:lasdamas_map}3D projection of the galaxy number density field corresponding to one of the Las Damas mocks. The colour palette used in the projection is the same as in Figures~\ref{fig:data_map} and~\ref{fig:sims_maps}. Notice the similarity of the clustering in this case and that of the real data shown in Figure~\ref{fig:data_map}.}
\end{center}
\end{figure}

For each of the mocks, we compute the galaxy density field $\Dgi$ using the same method described in section~\ref{subsec:sdss_data} for the SDSS data, including the use of the angular and radial selection functions, and keeping only cells with completeness $c_i \geq 0.8$.
For consistency, we use here the same cosmological model used to create the simulations (which is slightly different from the \emph{Planck-2015} model we use elsewhere). 
This results in a slightly different number of valid cells ($N = 589$) with respect to the SDSS data catalogue. 
In section~\ref{subsec:lasdamas_results} we use the Las Damas mocks to test the reliability of our method with respect to the cosmological model used here. In Figure~\ref{fig:lasdamas_map} we show a 3D projection of the galaxy density field for one of the Las Damas mocks. By comparing to Figure~\ref{fig:data_map} we see that the clustering of the mock seems very similar to that of the real data.

\section{Tests of the method}
\label{sec:tests}

We tested the N-pdf method to constrain $b$, $\sigma_8$ presented in section~\ref{sec:model} using the two sets of simulated galaxy density fields presented in sections~\ref{subsec:logn_simus} and \ref{subsec:lasdamas_simus}. First, in section~\ref{subsec:logn_simus} we show the results of applying our method to the lognormal simulations generated using this same model for the density field. This is a way to assess the internal consistency of the method. Then, in section~\ref{subsec:lasdamas_results}, we analyse the Las Damas mock catalogues. In this case, we can assess the reliability of the method in a realistic case when the basic assumptions (lognormal distribution of matter, linear biasing) are approximate but not exactly fulfilled.

\subsection{Application to lognormal simulations}
\label{subsec:logn_results}

We investigated the performance of the N-pdf method presented in section~\ref{sec:model} in the ideal case represented by the lognormal simulations described in section~\ref{subsec:logn_simus}.
In this way, we test the internal consistency of the method and, at the same time, we check whether the maximum-likelihood parameter estimator is, in fact, unbiased, and we explore how the estimator sensitivity depends on the actual bias factor.
We apply the method to the 100 simulations, for galaxy density fields generated with four different values of the true bias: $b_{\rm true} = 0.5, 1.0, 1.5, 2.0$.

\begin{figure}
\includegraphics[width=0.5\textwidth]{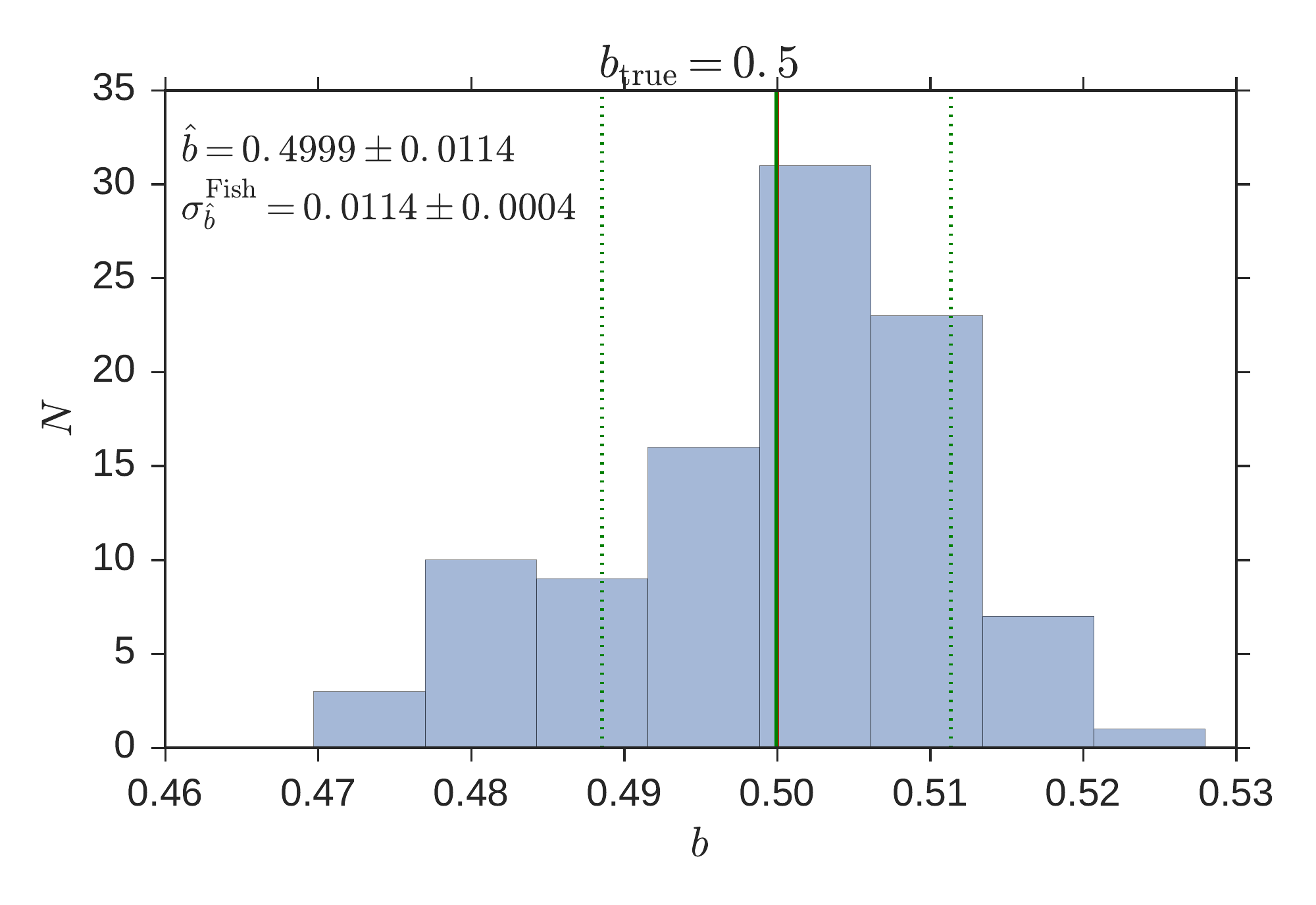}
\includegraphics[width=0.5\textwidth]{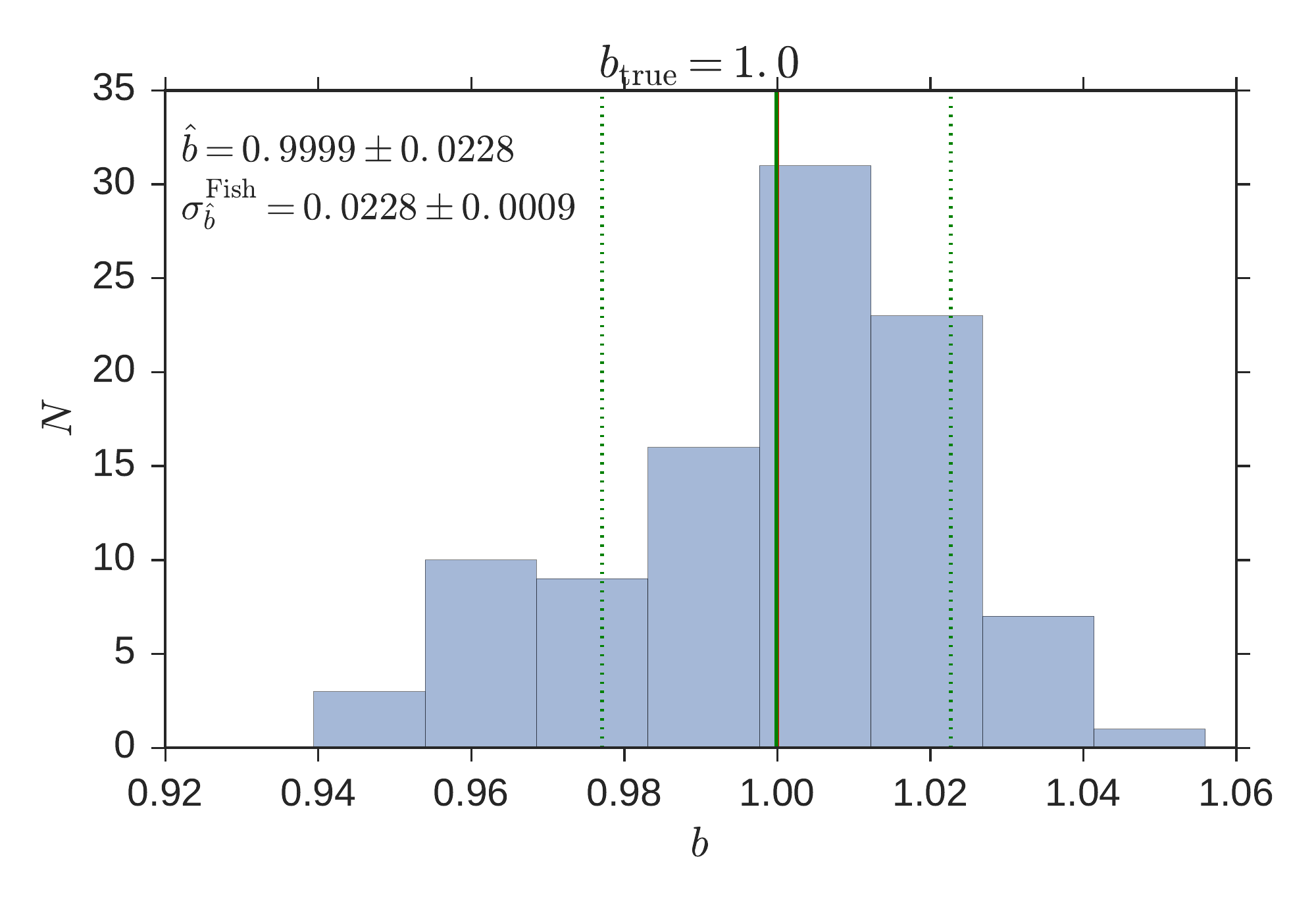}
\includegraphics[width=0.5\textwidth]{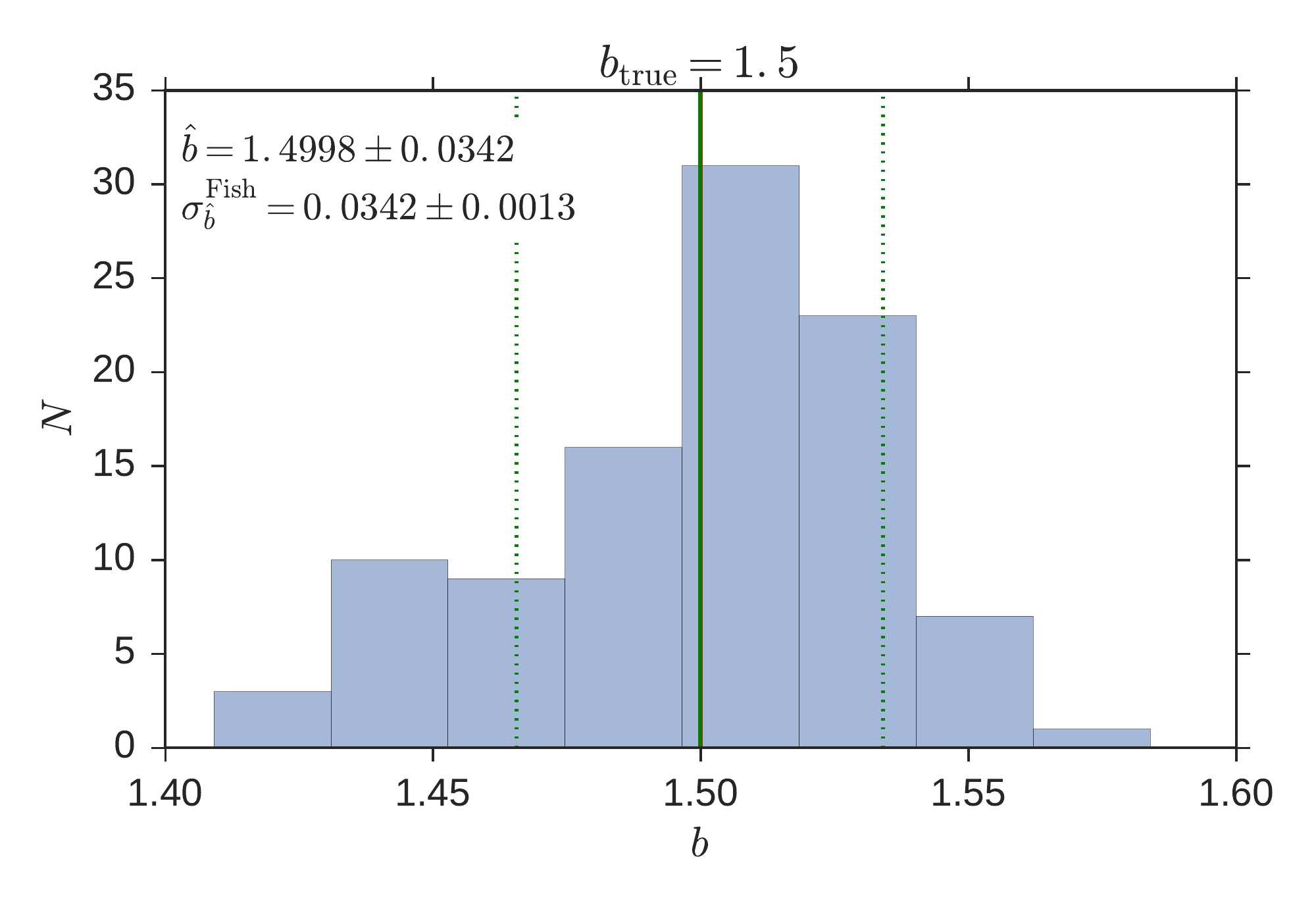}
\includegraphics[width=0.5\textwidth]{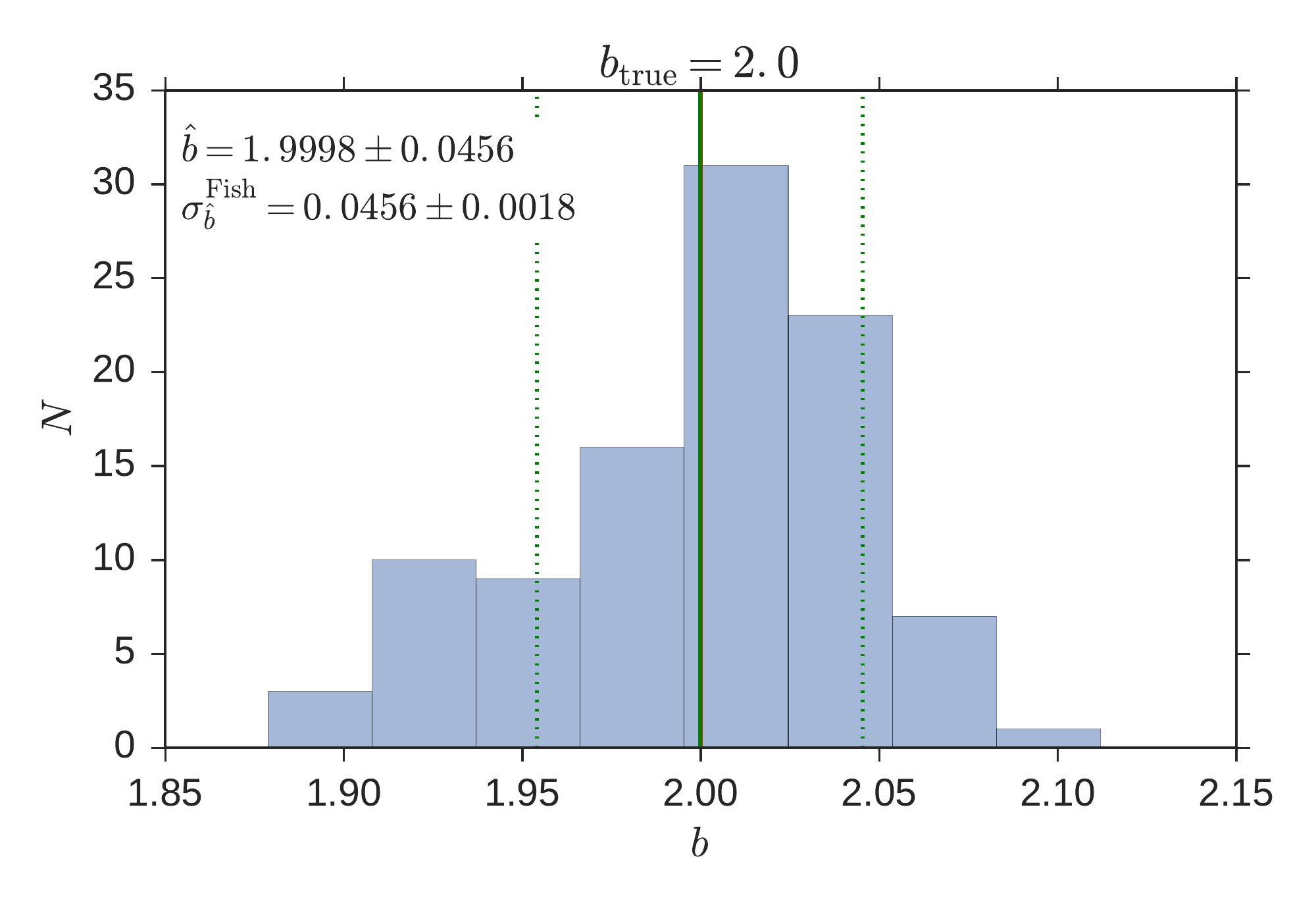}
\caption{\label{fig:sims_hist}
Distribution of the maximum-likelihood estimates of the bias $\bh$, for the 100 lognormal simulations, for the case in which the amplitude of the matter power spectrum is fixed. Each of the panel corresponds to a different value of the true bias in the simulations.
From left to right and top to bottom, these values are $b_{\rm true} = 0.5, 1.0, 1.5, 2.0$.
Input values are given as vertical red lines, whereas the mean $\bh$ over the simulations are represented by solid vertical green lines. The dashed green lines encompass the $1\sigma$ region around the mean.
We note in each case the mean $\bh$ and the mean uncertainty $\sigma_{\sh}$ obtained via the Fisher matrix.}
\end{figure}

First, we applied the method to the simulated fields assuming that the cosmological model was fully known, so that $\Cm$ was kept fixed to the true values of the matter correlation function, and we only fit for the galaxy bias $b$.
In Figure~\ref{fig:sims_hist} we present the histograms of the values of $\bh$ recovered for each realization. Each panel of the figure corresponds to a different value of the true bias.
In each panel, we list the mean value of our maximum-likelihood estimate $\bh$, together with its dispersion in the 100 simulations.
For each simulation, we also estimate the Fisher matrix uncertainty on the bias according to equation~(\ref{eq:post}). We also list in each panel the mean value and dispersion of the $\sigma_{\bh}$ obtained in this way.

We notice first that, as expected, the bias estimation is unbiased and that, second, the error associated to the parameter estimation is very well described by the Fisher matrix. Finally, the parameter dispersion increases as the bias increases, but the relative error in the bias determination (i.e., $\sigma_{\bh}/\bh$) is almost constant for all the cases: $\approx 2.3\%$.

\begin{figure}
\includegraphics[width=0.5\textwidth]{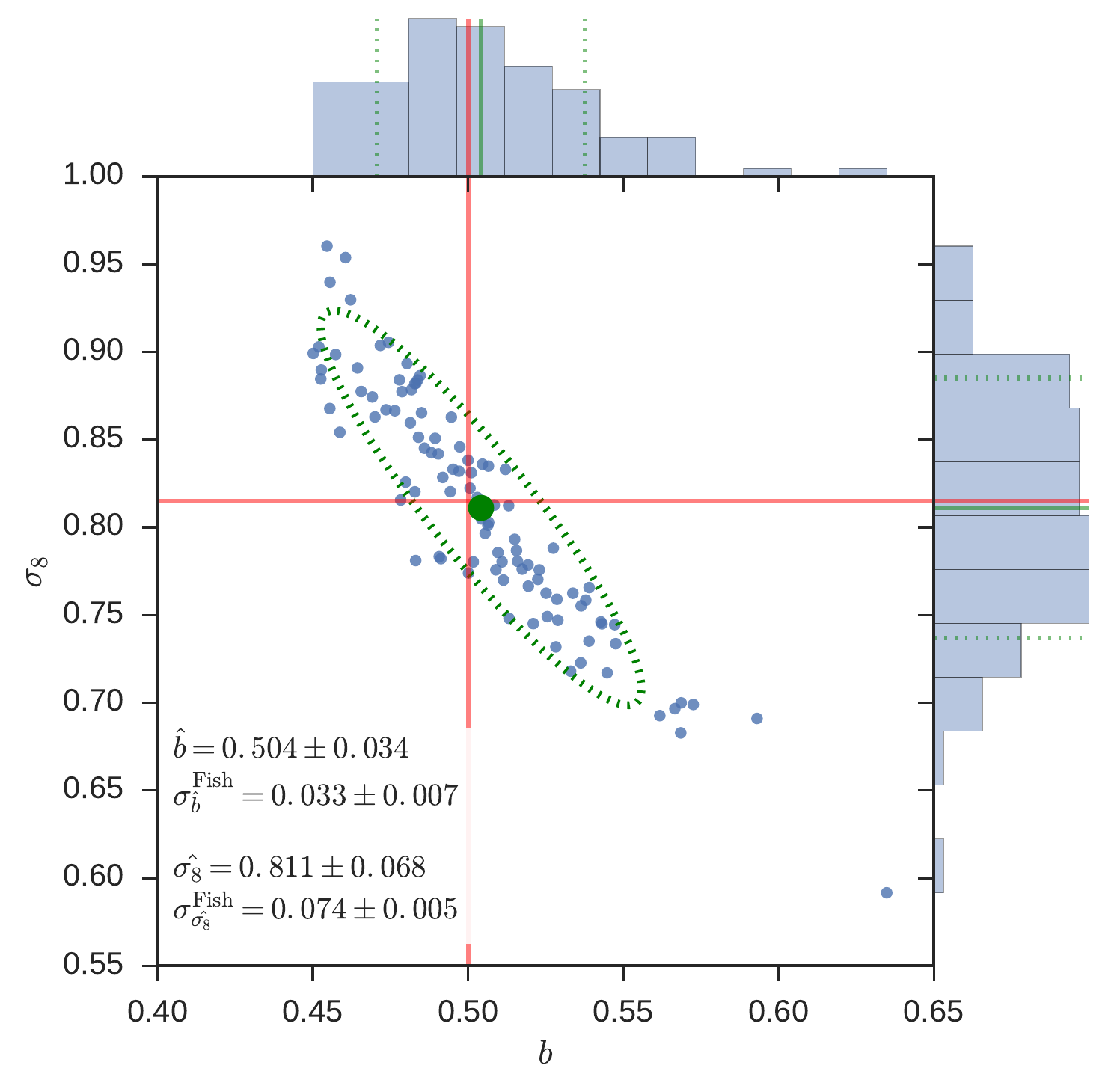}
\includegraphics[width=0.5\textwidth]{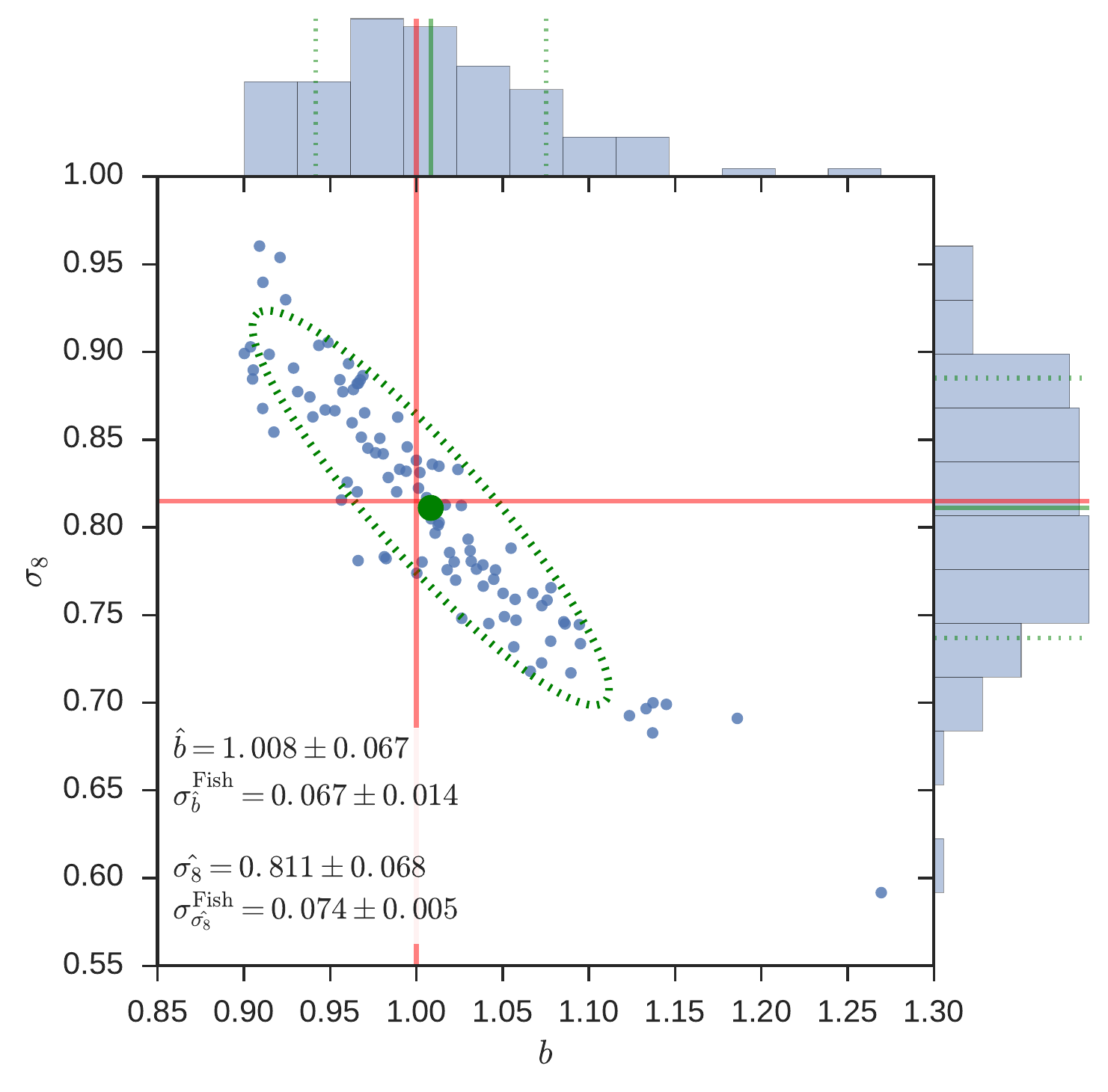}
\includegraphics[width=0.5\textwidth]{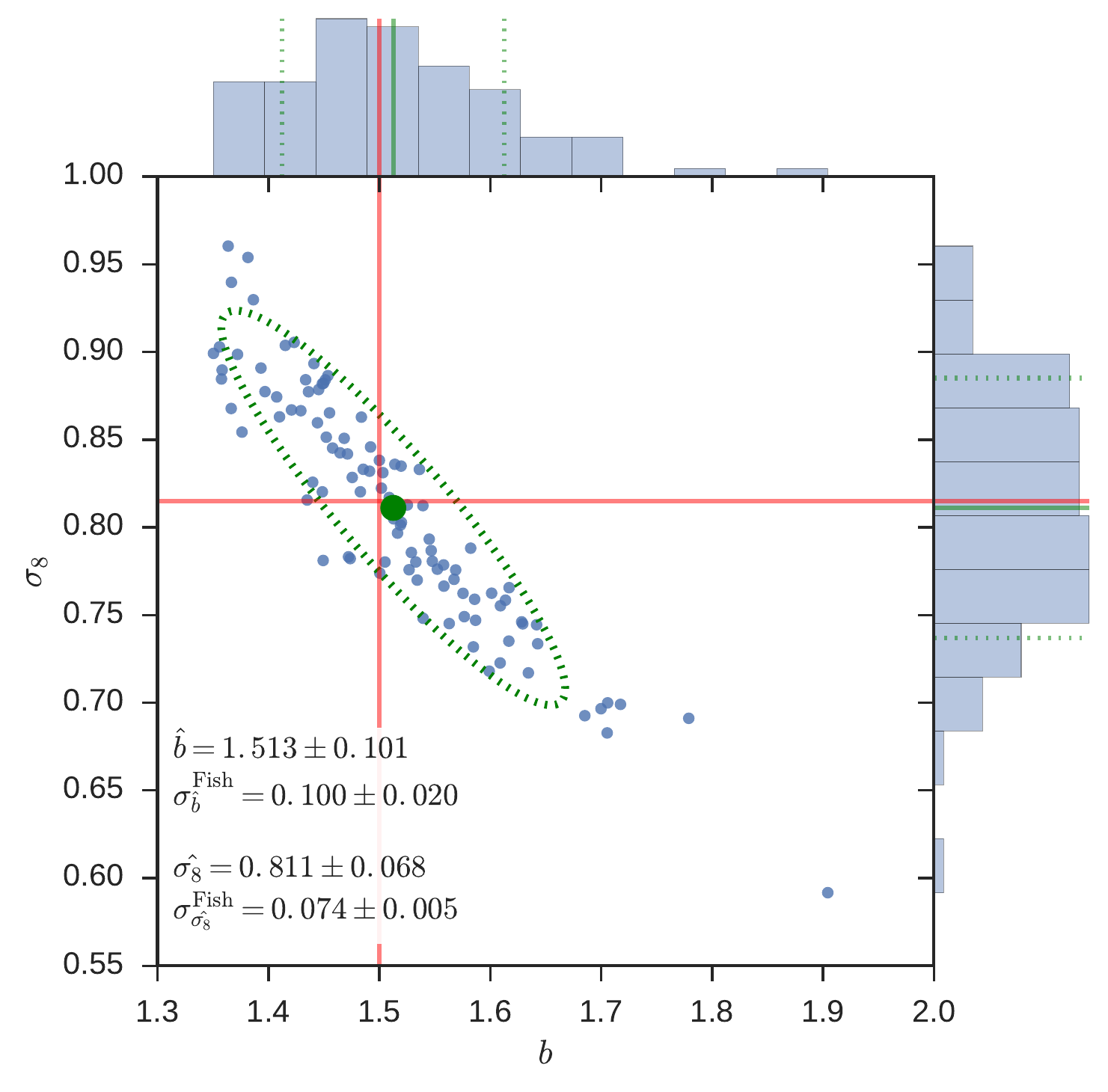}
\includegraphics[width=0.5\textwidth]{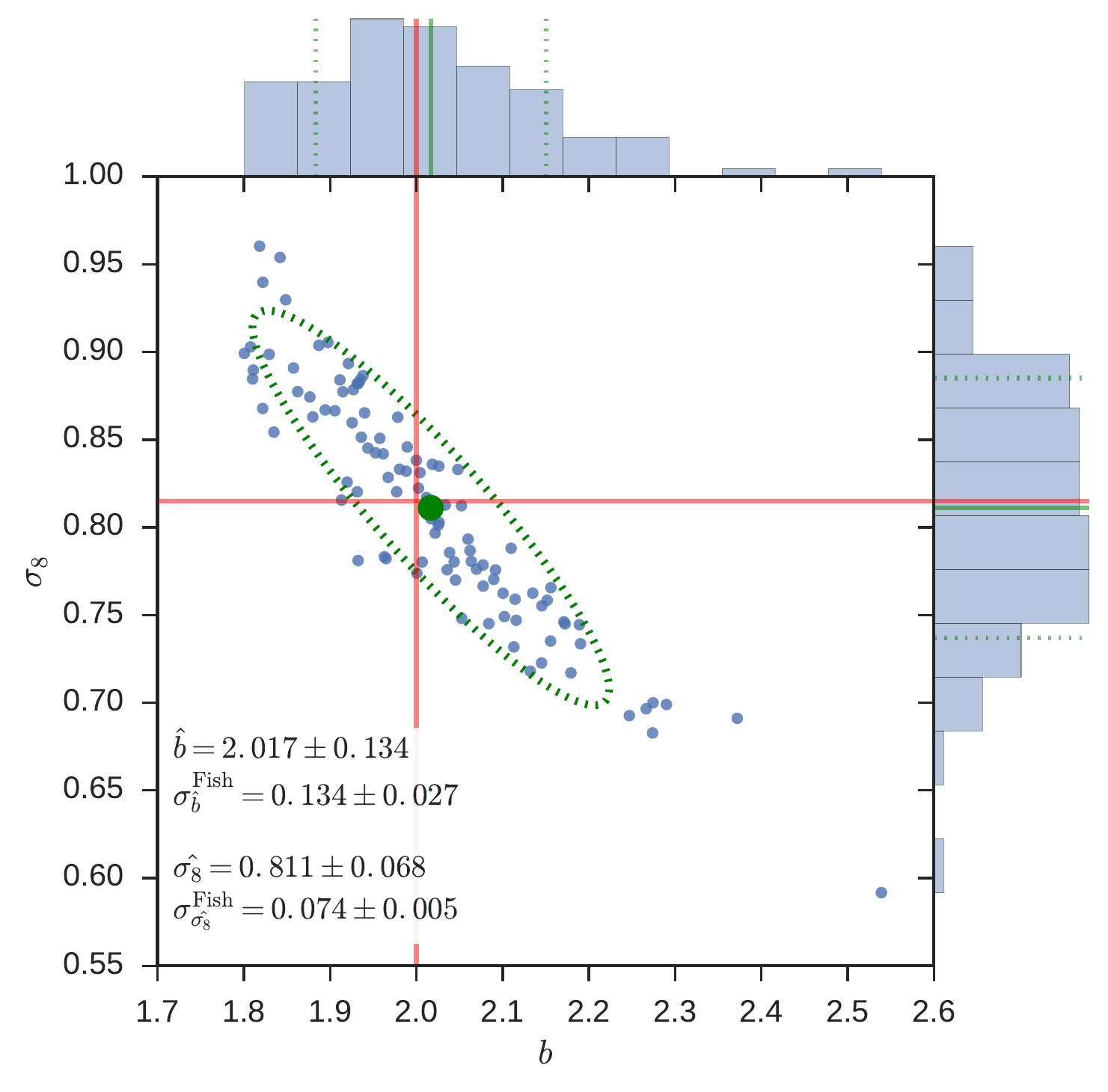}
\caption{\label{fig:sims_dist2d} 
Distribution of the maximum-likelihood estimates of the parameters $b$, $\sigma_8$, for the 100 lognormal simulations, for the case in which we allow both parameters to change.
From left to right and top to bottom, the panels correspond to different values of the true bias: $b_{\rm true} = 0.5, 1.0, 1.5, 2.0$.
Blue dots represent the individual estimates for each realization and the histograms at the top and right of each panel show the marginal distributions for each parameter.
Input values are given as red lines. The big green dot and solid lines correspond to the mean of the recovered parameters over the simulations. The dashed green lines in the histograms encompass the $1\sigma$ regions around the mean for each parameter. The green dashed ellipse represents the joint $1\sigma$ region corresponding to the mean covariance matrix from the Fisher matrix approach.
}
\end{figure}

As a second step, we perform the fit in both the galaxy bias $b$ and the power spectrum amplitude $\sigma_8$, as described by equations~(\ref{eq:cm_sigma}) and~(\ref{eq:like_bsigma}). We show the 2D distribution of the maximum-likelihood values ($\bh$, $\sh$) obtained from the 100 realisations in Figure~\ref{fig:sims_dist2d}, together with the marginal distributions of each of the two parameters. As before, each of the panels corresponds to a different true value of the bias. We also obtained in each case the parameter covariance matrix and the corresponding uncertainties using the Fisher matrix approach described in section~\ref{subsubsec:uncertainty}. In each panel, we show as a dashed contour the $1\sigma$ ellipse derived from the mean covariance matrix, and list the mean and dispersion of both the parameter estimations and their uncertainties.

We obtain, again, that the maximum-likelihood estimators of both $b$ and $\sigma_8$ are unbiased. For the case of the bias, we obtain again that the Fisher matrix estimation of the uncertainty matches the observed dispersion of $\bh$. As in the first case, this uncertainty scales with the bias, although it is significantly larger now, with $\sigma_{\bh}/\bh \approx 6.7\%$. This is understandable, as now there is additional freedom due to the value of $\sigma_8$ being allowed to change. Regarding the uncertainty on $\sigma_8$, we see that the Fisher matrix approach sligthly over-estimates it (by a $\approx 9\%$). Hence, we can take it as a conservative uncertainty estimate. We also observe that the error on $\sh$ does not depend on the (true) value of the bias.

From the analysis of the lognormal simulations, we can conclude that our method is consistent, in the sense that, when its assumptions are fullfilled, it provides unbiased estimates of $b$ and $\sigma_8$. Moreover, the Fisher matrix approach provides correct (or slightly conservative) estimates of the uncertainty.

\subsection{Application to Las Damas simulations}
\label{subsec:lasdamas_results}

To further test the reliability of the N-pdf method, and its applicability to the real distribution of galaxies in the Universe, we applied it to the Las Damas mocks described in section~\ref{subsec:lasdamas_simus}, which provide a test bench closely matching the galaxy distribution in the SDSS catalogue.
We first applied the method, both fitting only for the bias and for both the bias and the amplitude $\sigma_8$, using as fiducial model the true cosmological model used to create the simulations (see section~\ref{subsec:lasdamas_simus}). 
Regarding the input value of $b$ for the simulations, we know that they were built to match the projected correlation function of the real SDSS data. We therefore take as the `true' bias the value of $b = 1.20 \pm 0.01$ obtained by \cite{zeh10a} for the corresponding SDSS catalogue\footnote{The analysis in \cite{zeh10a} uses the same Las Damas cosmology, so we can use this value directly in our comparison} by a HOD fit to $w_p(r_p)$. We note that other fitting methods in that same work gave compatible values of the bias with significantly larger error.

\begin{figure}
\centering
\includegraphics[width=0.48\textwidth]{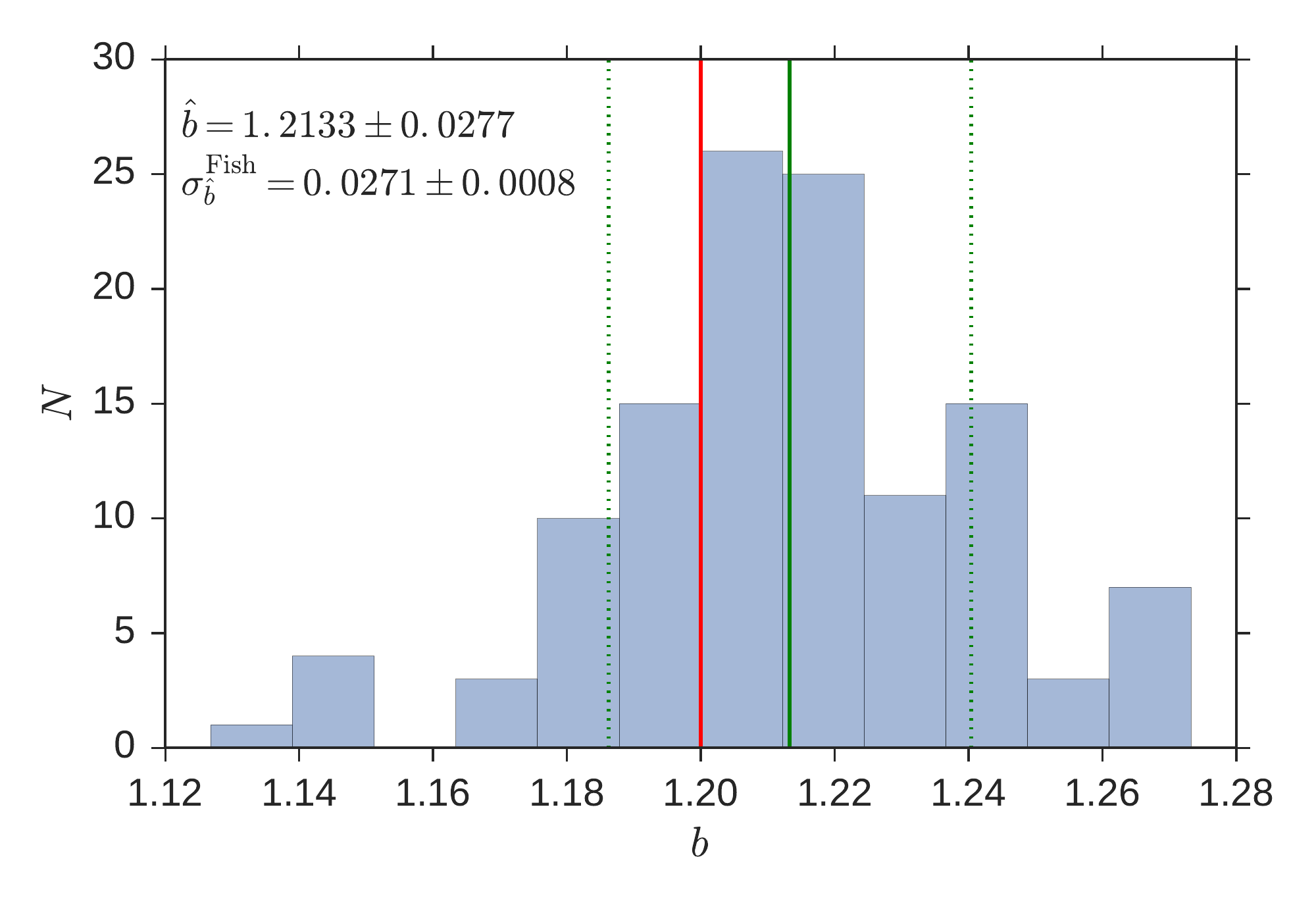}
\includegraphics[width=0.48\textwidth]{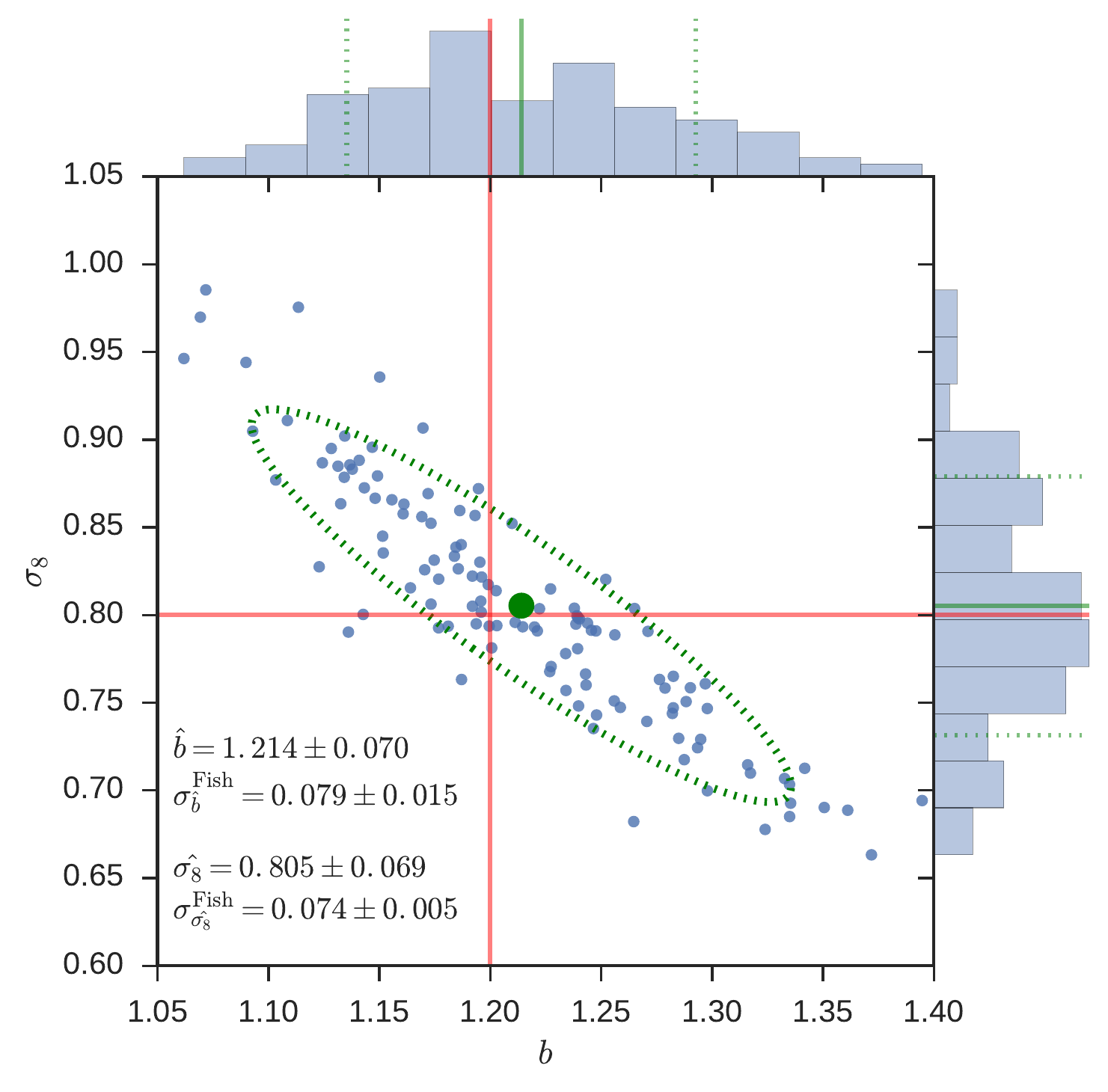}\\
\caption{\label{fig:lasdamas_hist} 
(Left) Distribution of the maximum-likelihood estimates of the bias, for the 120 realizations of the Las Damas mocks, for the case in which $\sigma_8$ is fixed. The vertical lines have the same meaning as in Figure~\ref{fig:sims_hist}.
(Right) Distribution of the maximum-likelihood joint estimates of $b$ and $\sigma_8$, for these same mocks, for the case in which we fit for both parameters. The meaning of the different symbols and lines is the same as in Figure~\ref{fig:sims_dist2d}.
}
\end{figure}

The results of the application of the N-pdf method to the Las Damas mocks are shown in Figure~\ref{fig:lasdamas_hist}. The left panel shows the distribution of maximum-likelihood estimates of the bias, $\bh$, when we keep all the cosmological parameters fixed to their true values. 
In this case, we obtain the estimates of the bias $\bh = 1.2133 \pm 0.0277$. 
The uncertainties derived via the Fisher matrix are $\sigma_{\sh}^{\rm Fish} = 0.0271 \pm 0.0008$, which provide again a very good estimate of the true dispersion of the recovered $\bh$. The relative error on the estimated bias, $\sigma_{\bh}/\bh = 2.3\%$ is remarkably close to the one obtained for the lognormal simulations in the previous section. The recovered bias is compatible, within the errors, with the `true' value. We note that we do not expect here a perfect agreement, as the input bias was fixed using a different statistic ($w_p(r_p)$) and range of scales.

The right panel of Figure~\ref{fig:lasdamas_hist} shows the 2D distribution of the estimates $\bh$, $\sh$ when we also allow the power spectrum amplitude to vary. In this case, we obtain again results very similar to the ones obtained for the lognormal simulations. We obtain a relative error of the bias $\sigma_{\bh}/\bh = 5.8\%$, with the true dispersion being well recovered by the Fisher matrix estimate. We recover the amplitude of the power spectrum as $\sh = 0.805 \pm 0.069$, in perfect agreement with the true value of $\sigma_8 = 0.8$. The Fisher matrix estimate of the uncertainty, $\sigma_{\sh}^{\rm Fish} = 0.074 \pm 0.005$, slightly overestimates the error on $\sh$.

Overall, we obtain that the results for the Las Damas mocks are remarkably similar to those obtained for the case of the lognormal simulations and that the N-pdf method provides also in this case unbiased estimations of $b$ and $\sigma_8$.
This is an indication that the basic assumptions of the method (lognormal distribution of the matter density field and linear biasing) are good enough approximations for our purposes in realistic distributions of galaxies, even if we now that they are not exact.
In this way, we have validated the method, and we can confidently use it in real galaxy catalogues, as we do below.

\subsubsection{Robustness with respect to the fiducial cosmological model}
\label{subsubsec:cosmol_model}

We also used the Las Damas mocks to assess how our results may depend on the assumed fiducial cosmology. We repeated the analysis for the 120 Las Damas mocks using three different cosmological models (in addition to the original one used above). 
We first tested the effect of changing the fiducial value of $\sigma_8$ alone. In this case, we kept all the rest of the cosmological parameters fixed to the true Las Damas cosmology, while changing the amplitude of the fiducial power spectrum. This results in a change of the covariance matrix $\Cm$ used in the analysis (see equation~\ref{eq:cm_sigma}). We studied two cases, $\sigma_8^{\rm fid} = 0.7, 0.9$, which encompass in excess the range of plausible values given our present knowledge \cite[see, e.g.,][]{PlanckCollaboration2015b}. 

As a second test, we considered an overall change in all the cosmological parameters, and used the set of values compatible with the \emph{Planck-2015} results that we use to analyse the real data. In addition to $\sigma_8$, the main parameter potentially affecting our analysis is $\Omega_m$, which changes from $\Omega_m=0.25$ in the Las Damas cosmology to $\Omega_m=0.308$ in the \emph{Planck-2015} model. This change in parameters not only affects the covariance matrix $\Cm$ used for the parameter estimation, but also the density field $\Dg$ itself (as we change the relation between redshifts and distances). For this test, therefore, we repeat the estimation of $\Dg$ with the new cosmology, following section~\ref{subsec:lasdamas_simus}.

\begin{figure}
\centering
\includegraphics[width=0.5\textwidth]{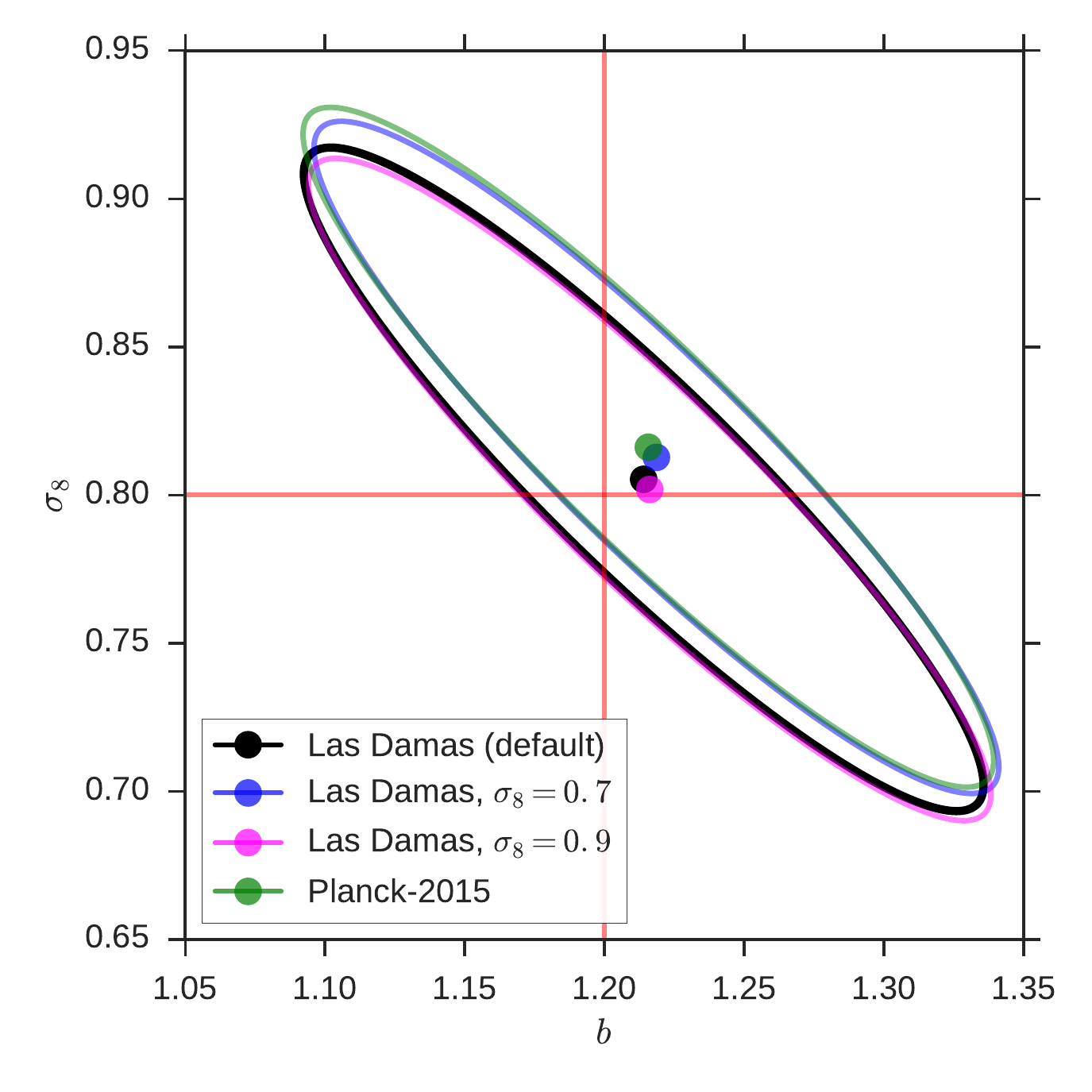}
\caption{\label{fig:lasdamas_cosmos} 
Average maximum-likelihood constraints on $b$, $\sigma_8$, for the 120 Las Damas mock realisations, when assuming different fiducial cosmological models in the analysis. The black point and contour correpond to the true Las Damas cosmology and are the same as shown in the right panel of Figure~\ref{fig:lasdamas_hist}. The blue and magenta colours correspond to the case in which we use the same Las Damas cosmology, except for the fiducial value of $\sigma_8$. The green colour corresponds to the analysis performed assuming the \emph{Planck-2015} cosmological parameters from \cite{PlanckCollaboration2015b}.}
\end{figure}

We summarize our results (for the case in which we fit for both $b$ and $\sigma_8$) in Figure~\ref{fig:lasdamas_cosmos} and Table~\ref{tbl:cosmopars}.
We see that, in all cases, the changes in the estimations of $\bh$, $\sh$ are much smaller than the uncertainties. In the case of the galaxy bias, the maximum relative change with respect to the default analysis is $0.4\%$, while the relative error of the measurement is $5.8\%$. Regarding the recovered value of $\sigma_8$, this relative change is $1.4\%$, compared to a relative error of $8.6\%$. The Fisher matrix estimation of the covariance matrix remains essentially unchanged for the different tested cosmologies.
We can conclude, therefore,  that our method is robust with respect to plausible changes in the fiducial cosmological model used in the analysis.

\begin{table}
\begin{center}
\begin{tabular}{|l||c|c|c|c|}
\hline
Fiducial cosmology   & $\bh$ & $\sigma_{\bh}^{\rm Fish}$ &  $\sh$ & $\sigma_{\sh}^{\rm Fish}$ \\ \hline
Las Damas & $1.214 \pm 0.070$  & $0.079 \pm 0.015$ & $0.805 \pm 0.069$ & $0.074 \pm 0.005$ \\
Las Damas, $\sigma_8 = 0.7$ & $1.219 \pm 0.071$ & $0.079 \pm 0.015$ & $0.813 \pm 0.070$ & $0.075 \pm 0.005$ \\
Las Damas, $\sigma_8 = 0.9$ & $1.216 \pm 0.071$ & $0.079 \pm 0.015$ & $0.802 \pm 0.069$ & $0.074 \pm 0.005$ \\
\emph{Planck-2015} & $1.216 \pm 0.072$ & $0.080 \pm 0.016$ & $0.816 \pm 0.069$ & $0.075 \pm 0.005$ \\ \hline
\end{tabular}
\caption{\label{tbl:cosmopars} 
Results obtained for the maximum-likelihood estimates of the parameters ($\bh$, $\sh$) and their Fisher-matrix uncertainties ($\sigma_{\bh}$, $\sigma_{\sh}$), in the 120 Las Damas mocks, when using different fiducial cosmologies in the analysis, as explained in the text.
In each case, we give the mean value and dispersion over the 120 realisations.
}
\end{center}
\end{table}

\section{Results for SDSS data}
\label{sec:sdss}

\begin{figure}
\begin{center}
\includegraphics[width=0.7\textwidth]{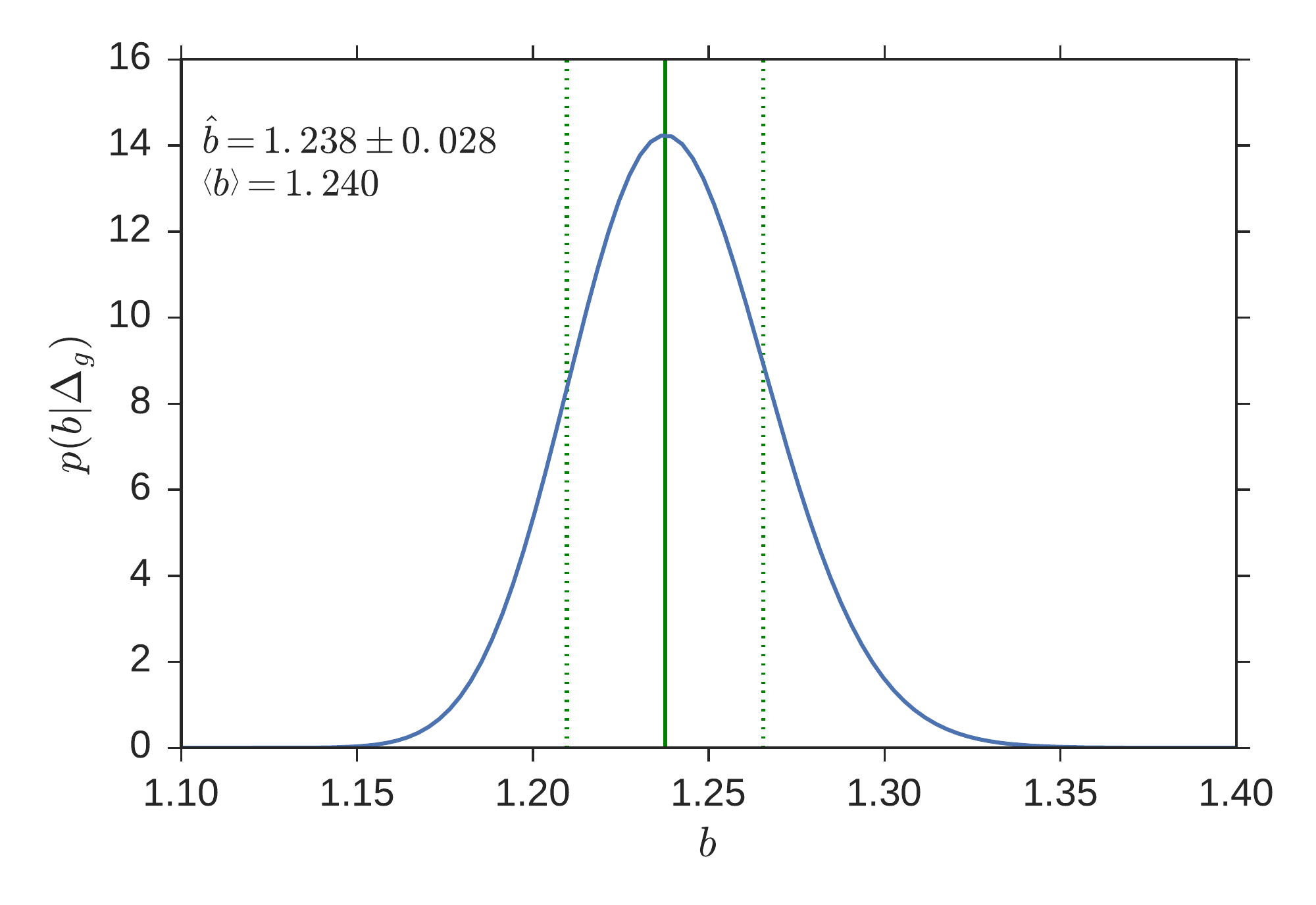}
\caption{\label{fig:data_pdf}
Posterior probability distribution of the bias parameter obtained by analysing the SDSS catalogue described in section~\ref{subsec:sdss_data}, for the case in which $\sigma_8$ is fixed at its fiducial value, $\sigma_8^{\rm fid} = 0.8149$. The solid vertical line marks the maximum-likelihood estimate $\bh$, and the dashed lines correspond to the $1\sigma$ interval estimated from the Fisher matrix analysis. 
}
\end{center}
\end{figure}

We used the N-pdf method presented in section~\ref{sec:model} to estimate the galaxy bias of the $M_r < -20$ sample and $\sigma_8$ using the galaxy density field $\Dg$ obtained from SDSS data as described in section~\ref{subsec:sdss_data}. We first consider the case in which $\sigma_8$ is fixed at its fiducial value (here, $\sigma_8^{\rm fid} = 0.8149$), and fit only for the galaxy bias. The posterior pdf of the bias in this case is shown in Figure~\ref{fig:data_pdf}, and the maximum-likelihood estimate obtained from the analysis is $\bh = 1.238 \pm 0.028$. The mean bias is $\bb = 1.240$.

This result can be compared to the bias estimated for the same SDSS sample by \cite{zeh10a}, using the projected correlation function $w_p(r_p)$. They obtained a value of $b = 1.20 \pm 0.01$ using HOD modelling, and $b = 1.17 \pm 0.07$ from the ratio of the galaxy and matter correlation functions in the range $r_p \in [4, 30]\, h^{-1}\,\mathrm{Mpc}$ (see their figure~7).\footnote{These measurements correspond to $b = 1.18 \pm 0.01$ and $b = 1.15 \pm 0.07$, respectively, when we make the transformation from the fiducial value of $\sigma_8$ used in \cite{zeh10a} to the one used here.}
There is a small discrepancy between these measurements and our results at only the $1-2\sigma$ level. 
This difference may just be due to the different methods used.
As the range of scales used is also different (in our case, the scales probed are larger than the size of the cells, and therefore $\geq 30 \,h^{-1}\,\mathrm{Mpc}$), this could also be related to a scale dependent bias \cite{Fang1998f,Blanton1999b,Coles2007b,Desjacques2010c}.

\begin{figure}
\begin{center}
\includegraphics[width=0.7\textwidth]{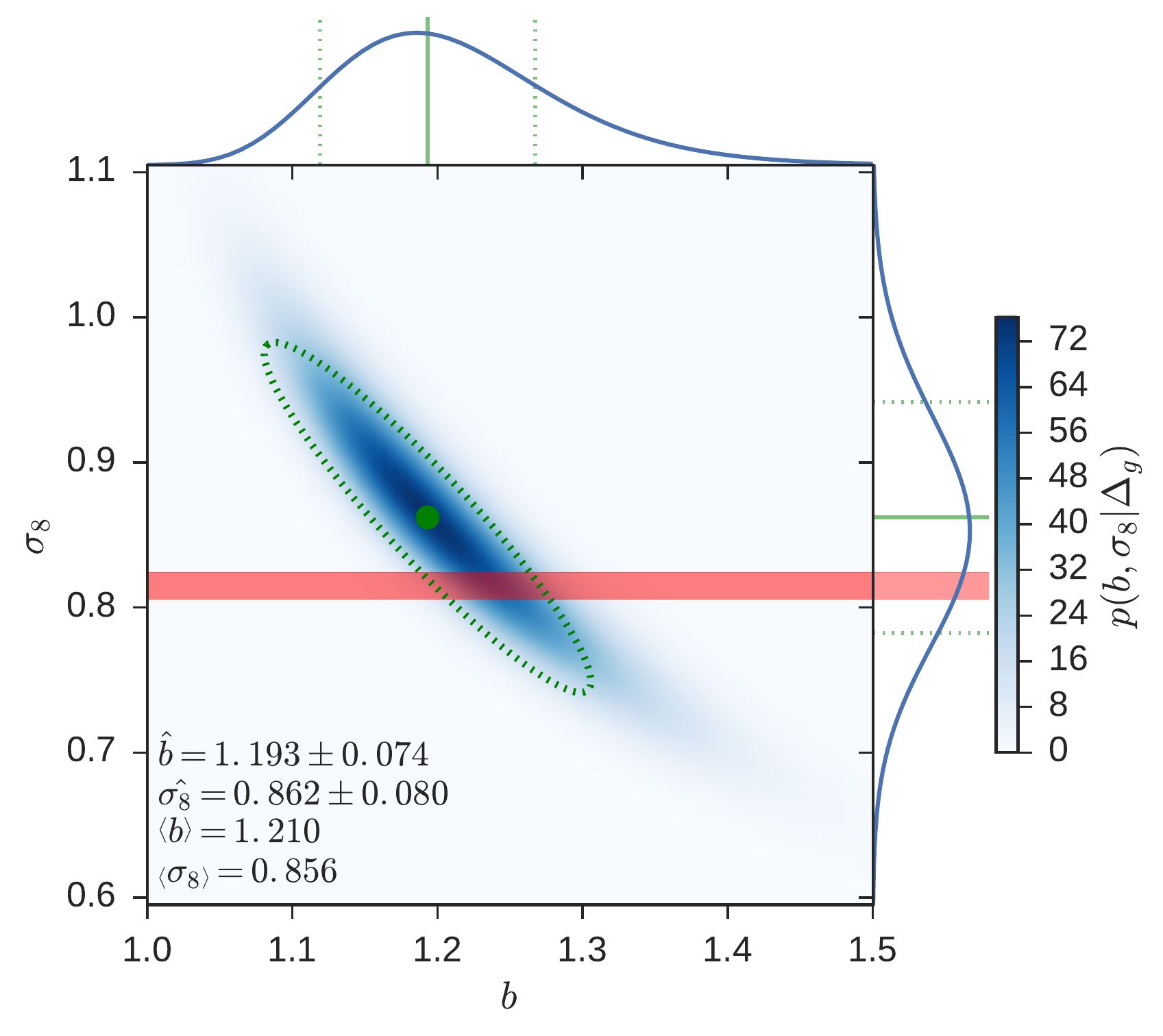}
\caption{\label{fig:data_pdf_2d} 
Joint posterior probability distribution of the two parameters $b$, $\sigma_8$ obtained by analysing the SDSS catalogue. The top and right sub-panels correspond to the marginalized probability distribution for each of the parameters separately.
The green dot and dashed contour show the maximum-likelihood estimate of the parameters, and the $1\sigma$ confidence region obtained from the Fisher matrix analysis.
The red horizontal band corresponds to the $1\sigma$ region from the measurement of $\sigma_8$ by \emph{Planck-2015} \cite{PlanckCollaboration2015b}. We see that our constraints on $\sigma_8$ are fully compatible with the ones coming from CMB analysis.}
\end{center}
\end{figure}

In Figure~\ref{fig:data_pdf_2d} we show the 2D joint posterior pdf for $b$ and $\sigma_8$ for the case in which we fit for both parameters. The marginal posterior distributions are also shown at the top and right sides of the figure, respectively.
The figure shows that we can indeed disentangle the constraints on the matter power spectrum from those on galaxy biasing.
The corresponding maximum-likelihood estimates are $\bh = 1.193 \pm 0.074$, $\sh = 0.862 \pm 0.080$. From the Fisher matrix estimate of the covariance matrix, we derive the correlation coefficient between the parameters as $r = \frac{C_{12}}{\sigma_{\bh}\sigma_{\sh}} = -0.937$

The bias estimate obtained from this analysis has a larger uncertainty than in the previous case (when $\sigma_8$ was fixed). However, it has the advantage of being an `absolute' measurement, in the sense that it does not depend on choosing a particular value of $\sigma_8$ as is needed when using the usual 2-point clustering statistics.

From this analysis we also measure the amplitude of the matter power spectrum at the redshift of the survey ($z_{\rm med} = 0.083$), which we express in terms of $\sigma_8$. As shown in Figure~\ref{fig:data_pdf_2d}, our measurement is in very good agreement with the value determined from \emph{Planck-2015} data, $\sigma_8 = 0.8149 \pm 0.0093$ \cite{PlanckCollaboration2015b}. We measure $\sigma_8$ with a precision of $9\%$, which is far from the $\approx 1\%$ precision that can be achieved from CMB measurements. However, this measurement can be complementary as it is done using a different method at a very different redshift. It can be therefore used to test the consistency of the cosmological model, in a similar way as other low redshift estimates of $\sigma_8$, e.g. via lensing \cite{hoe02,jee13}.

\subsection{Model selection}
\label{subsec:modsel_results}

We have shown that the N-pdf method breaks the degeneracy between the bias and the power spectrum amplitude using the shape of the distribution: it is lognormal for the matter density field, but differs from it (as described by equation~\ref{eq:pdfg}) for the galaxy density field. We can assess how effective the method is to make this distinction using the model selection criteria described in section~\ref{subsec:modelsel}. 

We compare two models. The \emph{alternative} hypothesis $H_1$ is the model used elsewhere in the paper, in which both parameters ($b$, $\sigma_8$) are left free, and the \emph{null} hypothesis $H_0$ is a model in which the value of galaxy bias is kept fixed at $b \equiv 1$, and $\sigma_8$ is the only free parameter. Hypothesis $H_0$ then corresponds to the case in which the N-pdf of the \emph{galaxy} density field is lognormal (as that of matter), even if the clustering amplitude is allowed to change via $\sigma_8$. We note that bias measurements relying on two-point statistics as the projected correlation function are unable to distinguish between these two models.

The results obtained, for the three criteria $\aic$, $\bic$ and $\glrt$ are presented in Table~\ref{tbl:selection}. The best-fit parameters obtained for hypothesis $H_1$ were presented in the previous section, in the case of $H_0$ we obtain $\sh = 1.144 \pm 0.039$.
As the $\aic$ and $\bic$ are approximations to the Bayes factors, we can use here the usual Jeffreys scale \cite{Jeffreys1961a,Kass1995a} to assess the evidence of both models.
According to that scale, the evidence against $H_0$ is `strong' in the case of the $\aic$, and `substantial' for the $\bic$ (this difference comes from the fact that the $\bic$ penalizes more heavily the addition of extra parameters). The $\glrt$ also favours $H_1$ in a similar way. 
We can conclude from this test that the N-pdf of the galaxy density field does not follow a lognormal distribution, and it is better described by the biased model given by equation~(\ref{eq:pdfg}).

\begin{table}
\begin{center}
\begin{tabular}{|l|c|}
\hline
Criterion   &   \\ \hline
$\aic(H_0) - \aic(H_1)$ & $12.6$ \\
$\bic(H_0) - \bic(H_1)$ & $8.22$ \\
$\glrt$: $\log\left(\frac{L_1}{L_0}\right)$ at $\nu=1$ &  $7.29$\\
\hline
\end{tabular}
\caption{\label{tbl:selection}
Results of different model selection criteria (from top to bottom: \emph{Akaike information criterion}, \emph{Bayesian information criterion} and \emph{generalized likelihood ratio test}) applied to the \emph{null} hypothesis $H_0$ (bias is fixed at $b \equiv 1$, only $\sigma_8$ is allowed to vary), and the \emph{alternative} hypothesis $H_1$ (both bias $b$ and $\sigma_8$ are allowed to vary). The different criteria provide either `substantial' or `strong' evidence in favour of the biased model ($H_1$).
}
\end{center}
\end{table}

\section{Conclusions and discussions}
\label{sec:final}

We have presented a full description of the N-probability density function of the galaxy number density fluctuations. The method follows the common assumption
\citep[e.g.][]{col91a,kay01a} that dark matter density fluctuations follow a local non-linear transformation of the initial energy density perturbations. The N-pdf of
the galaxy number density fluctuations is given in terms of the galaxy bias parameter and the cold dark matter correlations, parameterized, in our case, by its
amplitude ($\sigma_8$).

The N-pdf provides us the most complete tool to perform statistical analyses, in particular parameter estimation and model selection. Regarding the former, optimal
parameter estimation can be performed via the maximum-likelihood, since the N-pdf of the galaxy number density field, seen as a function of the bias
and the $\sigma_8$ parameters, is nothing but the likelihood of the data (i.e., the galaxy number density realization) given these parameters. Even more, Bayesian
inference can be also performed if any suitable information about the bias parameter is available in the form of a \emph{prior}.

In relation to model selection, we have explored some well known criteria based on the likelihood (notice that, in the case of known priors, other approaches as
Bayesian evidence, could be followed): the \emph{Akaike information criterion}, the \emph{Bayesian information criterion}, and the \emph{generalized likelihood
ratio test} (GLRT).

The methodology has been tested with SDSS-like simulations, both ideal log-normal realizations and mocks derived from the Las Damas project, showing, in both
cases, that the maximum-likelihood estimates of the galaxy bias and the dark matter correlation amplitude are unbiased. We have applied our formalism to the 7th release of the
SDSS main sample \citep{aba08a}, for a volume-limited subset with magnitude $M_r < -20$. We obtain $\bh = 1.193 \pm 0.074$ and $\sh = 0.862 \pm 0.080$,
for galaxy number density fluctuations in cells of the size  of $30h^{-1}$Mpc. The $\bh$ and $\sh$ errors are obtained from the Fisher matrix. These are in good
agreement with alternative estimates, as the mean bias and amplitude derived from the N-pdf.

The three model selection criteria mentioned above show that the \emph{alternative} hypothesis ($H_1$ of a galaxy bias parameter $b$ given by the maximum-likelihood
estimator) is favoured with respect to a no-biasing scenario given by the \emph{null} hypothesis $H_0$ of $b\equiv 1$.

Finally, we want to remark that our model assumes a constant bias over space, however the formalism can be generalized for a spatially varying bias. This is a more
realistic situation, since it is well known that galaxy bias is scale-dependent and it evolves with redshift. This generalization is in progress, and will be addressed in a future paper.

\section*{Acknowledgements}
We would like to thank an anonymous referee for his/her comments that have improved the quality and readability of this paper.
We acknowledge partial financial support from the  Spanish Ministry for Economy and Competitiveness and FEDER funds through grants AYA2010-22111-C03-02, AYA2012-39475-C02-01 and AYA2013-48623-C2-2, and Generalitat Valenciana project PrometeoII 2014/060.
The authors acknowledge the computer resources, technical expertise and assistance provided by the Spanish Supercomputing Network (RES) node at Universidad de Cantabria. 
We thank the LasDamas collaboration for providing the mock catalogs that were used in this study.

Funding for the SDSS and SDSS-II has been provided by the Alfred P. Sloan Foundation, the Participating Institutions, the National Science Foundation, the U.S. Department of Energy, the National Aeronautics and Space Administration, the Japanese Monbukagakusho, the Max Planck Society, and the Higher Education Funding Council for England. The SDSS Web Site is http://www.sdss.org/.

The SDSS is managed by the Astrophysical Research Consortium for the Participating Institutions. The Participating Institutions are the American Museum of Natural History, Astrophysical Institute Potsdam, University of Basel, University of Cambridge, Case Western Reserve University, University of Chicago, Drexel University, Fermilab, the Institute for Advanced Study, the Japan Participation Group, Johns Hopkins University, the Joint Institute for Nuclear Astrophysics, the Kavli Institute for Particle Astrophysics and Cosmology, the Korean Scientist Group, the Chinese Academy of Sciences (LAMOST), Los Alamos National Laboratory, the Max-Planck-Institute for Astronomy (MPIA), the Max-Planck-Institute for Astrophysics (MPA), New Mexico State University, Ohio State University, University of Pittsburgh, University of Portsmouth, Princeton University, the United States Naval Observatory, and the University of Washington.


\bibliography{bias_biblio}

\end{document}